\documentclass[12pt]{article}
\pdfoutput=1

\usepackage{amsmath}
\usepackage{amsfonts}
\usepackage{amscd}
\usepackage{amsthm}
\usepackage{setspace}
\usepackage{amssymb}

\usepackage{graphicx}
\usepackage{authblk}
\usepackage{caption}
\usepackage{ytableau}
\usepackage{mathtools}

\def\Z{\mathbb{Z}}

\def\C{\mathbb{C}}
\def\P{\mathbb{P}}
\def\H{\mathbb{H}}

\def\dg{\dagger}

\begin{document}

\begin{titlepage}

\begin{flushright}
KEK-TH-2094 
\end{flushright}

\vskip 1cm

\begin{center}

{\bf \Large Unbroken $E_7\times E_7$ nongeometric heterotic strings,\\
\vspace{0.5cm}
 stable degenerations and \\
\vspace{0.5cm}
 enhanced gauge groups in F-theory duals}

\vskip 1.2cm

Yusuke Kimura$^1$ 
\vskip 0.4cm
{\it $^1$KEK Theory Center, Institute of Particle and Nuclear Studies, KEK, \\ 1-1 Oho, Tsukuba, Ibaraki 305-0801, Japan}
\vskip 0.4cm
E-mail: kimurayu@post.kek.jp

\vskip 1.5cm
\abstract{\par Eight-dimensional non-geometric heterotic strings with gauge algebra $\mathfrak{e}_8\mathfrak{e}_7$ were constructed by Malmendier and Morrison as heterotic duals of F-theory on K3 surfaces with $\Lambda^{1,1}\oplus E_8\oplus E_7$ lattice polarization. Clingher, Malmendier and Shaska extended these constructions to eight-dimensional non-geometric heterotic strings with gauge algebra $\mathfrak{e}_7\mathfrak{e}_7$ as heterotic duals of F-theory on $\Lambda^{1,1}\oplus E_7\oplus E_7$ lattice polarized K3 surfaces. In this study, we analyze the points in the moduli of non-geometric heterotic strings with gauge algebra $\mathfrak{e}_7\mathfrak{e}_7$, at which the non-Abelian gauge groups on the F-theory side are maximally enhanced. The gauge groups on the heterotic side do not allow for the perturbative interpretation at these points. We show that these theories can be described as deformations of the stable degenerations, as a result of coincident 7-branes on the F-theory side. From the heterotic viewpoint, this effect corresponds to the insertion of 5-branes. These effects can be used to understand nonperturbative aspects of nongeometric heterotic strings.
\par Additionally, we build a family of elliptic Calabi--Yau 3-folds by fibering elliptic K3 surfaces, which belong to the F-theory side of the moduli of non-geometric heterotic strings with gauge algebra $\mathfrak{e}_7\mathfrak{e}_7$, over $\P^1$. We find that highly enhanced gauge symmetries arise on F-theory on the built elliptic Calabi--Yau 3-folds.
}  

\end{center}
\end{titlepage}

\tableofcontents
\section{Introduction}
\par F-theory/heterotic duality \cite{Vaf, MV1, MV2, Sen, FMW} states that F-theory \cite{Vaf, MV1, MV2} compactification on an elliptic K3 fibered Calabi--Yau $(n+1)$-fold describes a theory physically equivalent to heterotic compactification\footnote{Recent progress of heterotic strings can be found, for example, in \cite{BHW0504, BHW0507, KM2009, ACGLP1307, AGS1402, OS1402, AT1405, OHS1509, COO1605, AOMSS1806, COOS1810}.} on an elliptic Calabi--Yau $n$-fold. Non-perturbative aspects of heterotic theory can be studied by utilizing this duality. F-theory/heterotic duality is strictly formulated when the stable degeneration limit\footnote{Stable degenerations in F-theory/heterotic duality have been studied recently, for example, in \cite{AHK, BKW, BKL, CGKPS, MizTan, KRES}.} \cite{FMW, AM} is taken on the F-theory side in which K3 fibers split into pairs of half K3 surfaces.
\par Recently, eight-dimensional non-geometric heterotic strings with unbroken $\mathfrak{e}_8\mathfrak{e}_7$ algebra were constructed by Malmendier and Morrison \cite{MM} by utilizing the F-theory/heterotic duality. The Narain space \cite{Narain}
\begin{equation}
D_{2,18} / O(\Lambda^{2,18})
\end{equation}
gives the moduli space of eight-dimensional heterotic strings, and the double cover of this space,
\begin{equation}
D_{2,18} / O^+ (\Lambda^{2,18}),
\end{equation}
is equivalent to the moduli space of F-theory on elliptic K3 surfaces with a section. This is the statement of F-theory/heterotic duality. Malmendier and Morrison considered F-theory compactifications on elliptic K3 surfaces with $H\oplus E_8\oplus E_7$ lattice polarization, namely elliptically fibered K3 surfaces with a type $II^*$ fiber and a type $III^*$ fiber with a global section, and they constructed the moduli of heterotic strings with unbroken $\mathfrak{e}_8\mathfrak{e}_7$ algebra as the heterotic duals of them on the 2-torus. The moduli space of the non-geometric heterotic strings with unbroken $\mathfrak{e}_8\mathfrak{e}_7$ algebra constructed in \cite{MM} is given by 
\begin{equation}
D_{2,3} / O^+(L^{2,3}).
\end{equation}
Here $L^{2,3}$ denotes the orthogonal complement of $H\oplus E_8\oplus E_7$ inside the K3 lattice $\Lambda_{K3}$, and the non-geometric heterotic strings constructed in \cite{MM} possess $O^+(L^{2,3})$-symmetry. Here $O^+(L^{2,3})$\footnote{The authors of \cite{MMS} discussed connections of K3 surfaces with lattice polarizations, non-geometric heterotic strings, and $O^+(\Lambda^{2,2})$-modular forms.} mixes the complex structure moduli, the K\"ahler moduli and the moduli of Wilson line values. Therefore, the resulting heterotic strings do not have a geometric interpretation\footnote{The authors of \cite{HMW} discussed non-geometric type II theories.}; for this reason, the resulting heterotic strings are called {\it non-geometric} heterotic strings. A single Wilson line expectation value is non-zero for the non-geometric heterotic strings with unbroken $\mathfrak{e}_8\mathfrak{e}_7$ gauge algebra as constructed in \cite{MM}. The mathematical results of Kumar \cite{Kumar} and Clingher and Doran \cite{ClingherDoran2011, ClingherDoran2012}, which gave the Weierstrass equations of elliptic K3 surfaces with a global section with $E_8 E_7$ singularity, the coefficients of which are expressed as Siegel modular forms of even weight, were used in their construction.
\par Clingher, Malmendier, and Shaska \cite{CMSE7} extended the construction of non-geometric heterotic strings by Malmendier and Morrison to non-geometric heterotic strings with unbroken $\mathfrak{e}_7\mathfrak{e}_7$ algebra. F-theory compactifications on elliptic K3 surfaces with $H\oplus E_7\oplus E_7$ lattice polarization, namely K3 surfaces with a global section with two type $III^*$ fibers, were considered, and eight-dimensional non-geometric heterotic strings on $T^2$ were obtained as the heterotic duals in their construction. The moduli of the resulting heterotic strings is parametrized by the space 
\begin{equation}
D_{2,4} / O^+(L^{2,4}).
\end{equation}
Here $D_{2,4}$ is the symmetric space of $O(2,4)$, namely, $D_{2,4}$ is defined as $O(2)\times O(4)\backslash O(2,4)$. The symmetric space $D_{2,4}$ is also referred to as the {\it bounded symmetric domain of type $IV$}. Here $L^{2,4}$ denotes the orthogonal complement of $H\oplus E_7\oplus E_7$ in the K3 lattice $\Lambda_{K3}$. The complex structure moduli, K\"ahler moduli, and the moduli of Wilson line expectation values are mixed under the symmetry $O^+(L^{2,4})$, thus the heterotic strings constructed in \cite{CMSE7} also do not have a geometric interpretation. Two Wilson line expectation values are non-trivial in non-geometric heterotic strings with unbroken $\mathfrak{e}_7\mathfrak{e}_7$ algebra. See also \cite{GJ1412, LMV1508, FGLMM1603, MS1609, GLMM1611, FM1708, FGLMM1712, Kimura1810, Plauschinn2018} for recent progress on non-geometric heterotic strings.

\par In this note, we analyze theories that correspond to the points in the moduli of eight-dimensional non-geometric heterotic strings on the 2-torus $T^2$ constructed in \cite{CMSE7}, at which the ranks of the non-Abelian gauge groups are enhanced to 18 on the F-theory side. These are the maximal enhancements of the non-Abelian gauge groups on the F-theory side. We mainly consider $E_8\times E_8$ heterotic strings, rather than $SO(32)$ heterotic strings. (However, we do consider some applications to $SO(32)$ heterotic strings.) As only up to an $E_8\times E_8 \times U(1)^4$ gauge group can arise in eight-dimensional $E_8 \times E_8$ heterotic strings compactified on the 2-torus \cite{Kimura1810}, a consideration of the ranks of the non-Abelian gauge groups reveals that the gauge groups of the dual heterotic theories of these F-theory models do not allow for a perturbative interpretation. These heterotic strings include the non-perturbative effects of 5-branes. We find that these theories can be described as the deformations of the heterotic strings from the stable degeneration limit, in which the F-theory/heterotic duality strictly holds, and that these deformations result from the coincident 7-branes on the F-theory side. In the heterotic language, the effect of coincident 7-branes corresponds to the presence of 5-branes. 
\par When the non-Abelian gauge groups on F-theory on an elliptic K3 surface are enhanced to rank 18, K3 surfaces become {\it extremal} K3 surfaces. A K3 surface is called attractive, when it has the Picard number $\rho=20$, which is the highest value for a complex K3 surface. When an attractive K3 surface has an elliptic fibration with a section with the singularity type of rank 18, the elliptic fibration is referred to as extremal. Owing to the classification result in \cite{SZ}, the complex structures of extremal K3 surfaces on which non-Abelian gauge groups on F-theory compactifications are enhanced to rank 18 in the moduli can be determined, and this enables us to deduce the Weierstrass equations of extremal K3 surfaces. By analyzing the deduced Weierstrass equations, we study F-theory compactifications and the non-geometric heterotic duals at these special points in the moduli. 
\par We also discuss applications to $SO(32)$ heterotic strings in this study. We deduce the Weierstrass equations of elliptic K3 surfaces appearing as the compactification spaces of the F-theory duals of some $SO(32)$ heterotic strings, which are obtained as the transformations of $\mathfrak{e}_7\mathfrak{e}_7$ non-geometric heterotic strings.
\par In addition, we consider fibering elliptic K3 surfaces that belong to the F-theory side of the moduli of eight-dimensional $\mathfrak{e}_7\mathfrak{e}_7$ non-geometric heterotic strings, over $\P^1$, to build elliptically fibered Calabi--Yau 3-folds with a global section. We study F-theory compactifications on the resulting elliptic Calabi--Yau 3-folds\footnote{Recent discussions of F-theory compactifications on elliptic Calabi--Yau 3-folds can be found, e.g., in \cite{GM, KT0906, KMT0911, KMT1008, MTmatter, GM2, BG1112, MT1201, MT1204, T1205, JT1605, MPT1610, MMP1711, HT1805, LLW1810, Kimura1810}. The authors of \cite{BM, AGW1612, GW1804} discussed F-theory on Calabi--Yau 3-folds with terminal singularities.}. We find that highly enhanced gauge groups arise in these compactifications. It is mainly local F-theory model buildings that have been discussed in recent studies \cite{DWmodel, BHV, BHV2, DWGUT}. However, the global aspects of the geometry need to be considered to discuss the issues of gravity. We investigate F-theory on elliptically fibered Calabi--Yau 3-folds from the global perspective in this study.
\par A similar organization can be found in \cite{Kimura1810}.
\par This note is structured as follows. In Section \ref{sec2}, we briefly review F-theory compactifications, and we also review attractive K3 surfaces and extremal K3 surfaces that are technically necessary to analyze special points in the moduli of eight-dimensional non-geometric heterotic strings and F-theory duals. We also review the construction of non-geometric heterotic strings with unbroken $\mathfrak{e}_7\mathfrak{e}_7$ algebra in \cite{CMSE7}. 
\par In Section \ref{sec3}, we discuss the special points in the eight-dimensional non-geometric heterotic moduli with unbroken $\mathfrak{e}_7\mathfrak{e}_7$ at which the ranks of the non-Abelian gauge symmetries on the F-theory side are enhanced to 18. The gauge groups in the heterotic strings which correspond to these points do not allow for the perturbative interpretations. We demonstrate that these theories can be seen as deformations of the stable degenerations as a result of the coincident 7-branes on the F-theory side. We also discuss applications to $SO(32)$ heterotic strings. We derive the Weierstrass equations of K3 elliptic fibrations appearing as the compactification spaces of the F-theory duals of some $SO(32)$ heterotic strings. We determine the gauge groups that arise on F-theory compactifications, including the global structures of the gauge groups.
\par We build elliptically fibered Calabi--Yau 3-folds in Section \ref{sec4} by fibering examples of elliptic K3 surfaces, which belong to the F-theory side of the moduli of eight-dimensional unbroken $\mathfrak{e}_7\mathfrak{e}_7$ non-geometric heterotic strings, over $\P^1$. We analyze F-theory compactifications on the resulting elliptic Calabi--Yau 3-folds. First, we consider the higher-dimensional analog of the construction of genus-one fibered K3 surfaces without a global section\footnote{Recent studies of F-theory compactifications on genus-one fibered spaces lacking a global section can be found in, for example, \cite{BM, MTsection, AGGK, KMOPR, GGK, MPTW, MPTW2, BGKintfiber, CDKPP, LMTW, K, K2, KCY4, CGP, Kdisc, Kimura1801, AGGO1801, Kimura1806, TasilectWeigand, CLLO, TasilectCL, HT}.} to build genus-one fibered Calabi--Yau 3-folds without a section. This construction ensures that the resulting 3-folds in fact satisfy the Calabi--Yau condition. Similar constructions of genus-one fibered Calabi--Yau 4-folds without a section using double covers can be found in \cite{KCY4}. Taking the Jacobian fibration\footnote{\cite{Cas} discussed the Jacobians of elliptic curves.} of the resulting genus-one fibered Calabi--Yau 3-folds yields elliptically fibered Calabi--Yau 3-folds with a global section. K3 fibers of these elliptic Calabi--Yau 3-folds belong to the F-theory side of the moduli of eight-dimensional non-geometric heterotic strings with unbroken $\mathfrak{e}_7\mathfrak{e}_7$ algebra. Therefore, the obtained elliptic Calabi--Yau 3-folds can be seen as the fibering of such K3 surfaces over the base curve $\P^1$. We deduce the gauge groups on F-theory compactifications on the elliptic Calabi--Yau 3-folds, and we find that some specific models do not have a $U(1)$ gauge field. We determine the Mordell--Weil groups of some models, and we obtain the global structures of the gauge groups of these models. We also deduce candidate matter spectra on F-theory on the constructed elliptically fibered Calabi--Yau 3-folds that satisfy the six-dimensional anomaly cancellation condition. We determine these candidate matter spectra directly from the global defining equations of the elliptically fibered Calabi--Yau 3-folds. We state our concluding remarks in Section \ref{sec5}.

\section{Review of non-geometric heterotic strings with unbroken $\mathfrak{e}_7\mathfrak{e}_7$ algebra, F-theory, and extremal K3 surfaces}
\label{sec2}	

\subsection{Review of F-theory compactifications}
\label{ssec2.1}
We briefly review F-theory compactifications on elliptic K3 surfaces. A similar review can be found in \cite{Kimura1810}. F-theory is compactified on spaces that admit a genus-one fibration. The complex structure of the genus-one fiber is identified with the axio-dilaton in F-theory compactification. This formulation allows the axio-dilaton to have $SL(2,\Z)$ monodromy. Genus-one fibrations do not necessarily admit a global section; there are situations in which they have a global section, and those in which they do not. F-theory compactifications on elliptic fibrations with a global section have been investigated in recent studies, for example, in \cite{MorrisonPark, MPW, BGK, BMPWsection, CKP, BGK1306, CGKP, CKP1307, CKPS, AL, EKY1410, LSW, CKPT, CGKPS, MP2, BMW2017, CL2017, BMW1706, EKY1712, KimuraMizoguchi, EK1802, Kimura1802, LRW2018, EK1808, MizTani2018, CMPV1811, TT2019}. Although Calabi--Yau genus-one fibration lacking a global section cannot be expressed in the Weierstrass form, when the Jacobian fibration of it exists, the Jacobian fibration yields an elliptic fibration with a global section. Calabi--Yau genus-one fibration $Y$ and the Jacobian fibration $J(Y)$ have the identical types of the singular fibers, and they have the same discriminant loci. 
\par Genus-one fibers degenerate over the codimension 1 locus in the base space, and this locus is referred to as the discriminant locus. Such degenerate fibers are called the singular fibers. When genus-one fiber degenerates, it becomes either $\P^1$ with a single singularity, or a sum of smooth $\P^1$'s meeting in specific ways. The types of the singular fibers of genus-one fibered surfaces were classified by Kodaira \cite{Kod1, Kod2}. Methods to determine the singular fibers of elliptic surfaces can be found in \cite{Ner, Tate}. 
\par In F-theory compactifications on genus-one fibrations, the non-Abelian gauge groups that form on the 7-branes correspond to the singular fibers of genus-one fibrations \cite{MV2, BIKMSV}. The correspondences of the singular fibers and the singularity types of the compactification spaces are shown in Table \ref{tablemonodromy in 2.1} below. The corresponding monodromies and j-invariants of the singular fibers are also presented in the table. 

\begingroup
\renewcommand{\arraystretch}{1.1}
\begin{table}[htb]
  \begin{tabular}{|c|c|r|c|c|} \hline
Fiber type & J-invariant & Monodromy  & Order of Monodromy & Singularity Type \\ \hline
$I^*_0$ & regular & $-\begin{pmatrix}
1 & 0 \\
0 & 1 \\
\end{pmatrix}$ & 2 & $D_4$\\ \hline
$I_m$ & $\infty$ & $\begin{pmatrix}
1 & m \\
0 & 1 \\
\end{pmatrix}$ & infinite & $A_{m-1}$\\
$I^*_m$ & $\infty$ & $-\begin{pmatrix}
1 & m \\
0 & 1 \\
\end{pmatrix}$ & infinite & $D_{m+4}$\\ \hline
$II$ & 0 & $\begin{pmatrix}
1 & 1 \\
-1 & 0 \\
\end{pmatrix}$ & 6 & none.\\
$II^*$ & 0 & $\begin{pmatrix}
0 & -1 \\
1 & 1 \\
\end{pmatrix}$ & 6 & $E_8$\\ \hline
$III$ & 1728 & $\begin{pmatrix}
0 & 1 \\
-1 & 0 \\
\end{pmatrix}$ & 4 & $A_1$\\
$III^*$ & 1728 & $\begin{pmatrix}
0 & -1 \\
1 & 0 \\
\end{pmatrix}$ & 4 & $E_7$\\ \hline
$IV$ & 0 & $\begin{pmatrix}
0 & 1 \\
-1 & -1 \\
\end{pmatrix}$ & 3 & $A_2$\\
$IV^*$ & 0 & $\begin{pmatrix}
-1 & -1 \\
1 & 0 \\
\end{pmatrix}$ & 3 & $E_6$\\ \hline
\end{tabular}
\caption{\label{tablemonodromy in 2.1}Monodromies, j-invariants and the corresponding types of the singularities of singular fibers. ``Regular'' for j-invariant of $I_0^*$ fiber means that j-invariant can take any finite value in $\C$ for $I_0^*$ fiber.}
\end{table}
\endgroup

The types of singular fibers of elliptic surfaces can be determined from the vanishing orders of the coefficients of the Weierstrass equations. The correspondences of the fiber types and the vanishing orders of the Weierstrass coefficients are  shown in Table \ref{tableWeierstrasscoefficients in 2.1}. 

\begingroup
\renewcommand{\arraystretch}{1.5}
\begin{table}[htb]
\begin{center}
  \begin{tabular}{|c|c|c|c|} \hline
Fiber Type & Ord($f$) & Ord($g$) & Ord($\Delta$) \\ \hline
$I_0 $ & $\ge 0$ & $\ge 0$ & 0 \\ \hline
$I_n $  ($n\ge 1$) & 0 & 0 & $n$ \\ \hline
$II $ & $\ge 1$ & 1 & 2 \\ \hline
$III $ & 1 & $\ge 2$ & 3 \\ \hline
$IV $ & $\ge 2$ & 2 & 4 \\ \hline
$I_0^*$ & $\ge 2$ & 3 & 6 \\ \cline{2-3}
 & 2 & $\ge 3$ &  \\ \hline
$I^*_m$  ($m \ge 1$) & 2 & 3 & $m+6$ \\ \hline
$IV^*$ & $\ge 3$ & 4 & 8 \\ \hline
$III^*$ & 3 & $\ge 5$ & 9 \\ \hline
$II^*$ & $\ge 4$ & 5 & 10 \\ \hline   
\end{tabular}
\caption{\label{tableWeierstrasscoefficients in 2.1}List of the types of the singular fibers, and the corresponding vanishing orders of the coefficients, $f,g$, of the Weierstrass equation $y^2=x^3+f\, x+g$, and the orders of the discriminant, $\Delta$.}
\end{center}
\end{table}  
\endgroup 

\par When an elliptic fibration has a global section, the set of sections form a group, known as the Mordell--Weil group. The rank of the Mordell--Weil group gives the number of the $U(1)$ gauge fields in F-theory compactification on the elliptic fibration \cite{MV2}. 
\par The second integral cohomology group $H^2(S, \Z)$ of K3 surface $S$ includes the information of the geometry of the K3 surface. This group has the lattice structure, and it is called the K3 lattice, $\Lambda_{K3}$. The K3 lattice is unimodular, even lattice of signature (3,19), and it is isometric to the direct sum of two $E_8$'s and three hyperbolic planes \cite{Mil}
\begin{equation}
\Lambda_{K3} \cong E_8^2 \oplus H^3.
\end{equation}
The group of divisors (modulo algebraic equivalence) constitutes a sublattice inside the K3 lattice, called the N\'eron-Severi lattice $NS(S)$. When a K3 surface has an elliptic fibration with a global section, an elliptic fiber and a global section generate the hyperbolic plane $H$ inside the N\'eron-Severi lattice. K3 surface $S$ admitting an elliptic fibration with a section is equivalent to the condition that the N\'eron-Severi lattice $NS(S)$ contains the hyperbolic plane $H$ \cite{Kondoauto}. When an elliptic K3 surface has the singular fibers, the N\'eron-Severi lattice $NS(S)$ contains the $ADE$ lattices that correspond to the types of the singular fibers. For example, that a K3 surface $S$ is $H\oplus E_7\oplus E_7$-lattice polarized means that the N\'eron-Severi lattice $NS(S)$ includes the lattice $ H\oplus E_7\oplus E_7$, and this is equivalent to the condition that the K3 surface $S$ is elliptically fibered with a section, the singular fibers of which include two type $III^*$ fibers (or worse). K3 surfaces with $H\oplus E_7\oplus E_7$ lattice polarization are parametrized by the bounded symmetric domain of type $IV$, $D_{2,4}$, modded out by the symmetry of the orthogonal complement of the lattice $ H\oplus E_7\oplus E_7$ inside the K3 lattice $\Lambda_{K3}$:
\begin{equation}
\label{symmetry 2,4 in 2.1}
D_{2,4} / O^+(L^{2,4}).
\end{equation}
$L^{2,4}$ denotes the orthogonal complement of $ H\oplus E_7\oplus E_7$ inside the K3 lattice $\Lambda_{K3}$. Because the K3 lattice $\Lambda_{K3}$ is isometric to $E_8^2\oplus H^3$, $L^{2,4}$ can also be defined as the orthogonal complement of $E_7\oplus E_7$ in the lattice $E_8^2\oplus H^2$. 
\par By utilizing the F-theory/heterotic duality, eight-dimensional non-geometric heterotic strings with unbroken $\mathfrak{e}_7\mathfrak{e}_7$, the moduli space of which is equivalent to (\ref{symmetry 2,4 in 2.1}) were constructed in \cite{CMSE7}. In this note, we study the points in the moduli of such non-geometric heterotic strings at which the non-Abelian gauge symmetries are enhanced to rank 18 on the F-theory side. 

\subsection{Construction of $\mathfrak{e}_7\mathfrak{e}_7$ non-geometric heterotic strings by Clingher, Malmendier, and Shaska}
\label{ssec2.2}
We briefly review the construction of eight-dimensional non-geometric heterotic strings with unbroken $\mathfrak{e}_7\mathfrak{e}_7$ by Clingher, Malmendier and Shaska \cite{CMSE7}. 
\par As stated previously, the moduli of elliptic K3 surfaces with a global section with two $E_7$ singularities, namely the K3 surfaces with $H\oplus E_7\oplus E_7$ lattice polarization, are parameterized by the following space:
\begin{equation}
D_{2,4} / O^+(L^{2,4}).
\end{equation}
The bounded symmetric domain of type $IV$, $D_{2,4}$, is known to be isomorphic to ${\bf H}_2$ \cite{Matsumoto}:
\begin{equation}
\label{H2 isom in 2.2}
{\bf H_2} \cong D_{2,4}.
\end{equation}
${\bf H_2}$ is defined as 
\begin{equation}
{\bf H_2}:=\Big\{ \begin{pmatrix}
z_1 & z_2 \\
z_3 & z_4 \\
\end{pmatrix} \in M_2(\C) \hspace{2mm} | \hspace{2mm} 4 {\rm Im} \, z_1 {\rm Im} \, z_4 > |z_2-\overline{z_3}|^2 \hspace{2mm} {\rm and} \hspace{2mm} {\rm Im} \, z_4 > 0 \Big\}.
\end{equation}
As mentioned in \cite{CMSE7}, ${\bf H_2}$ is a generalization of the Siegel upper-half space $\H_2$ in the following sense:
\begin{equation}
\H_2 = \Big\{ \omega \in {\bf H_2} \hspace{2mm} | \hspace{2mm} \omega^t=\omega \Big\}.
\end{equation}
The modular group $\Gamma$ acting on ${\bf H_2}$ is defined as 
\begin{equation}
\Gamma  = \Big\{ G\in GL_4 (\Z[i]) \hspace{2mm} | \hspace{2mm} G^\dg \,  \begin{pmatrix}
0 & {\bf 1}_2 \\
-{\bf 1}_2 & 0 \\
\end{pmatrix} \, G = \begin{pmatrix}
0 & {\bf 1}_2 \\
-{\bf 1}_2 & 0 \\
\end{pmatrix} \Big\}.
\end{equation}
${\bf 1}_2$ denotes the 2 $\times$ 2 identity matrix. $\begin{pmatrix}
A & B \\
C & D \\
\end{pmatrix}$ in the modular group $\Gamma$ acts on $\omega \in {\bf H}_2$ as
\begin{equation}
\begin{pmatrix}
A & B \\
C & D \\
\end{pmatrix}\cdot \omega = (A\, \omega+B) (C\, \omega+D)^{-1}.
\end{equation}
There is an involution, ${\cal T}$, that acts on ${\bf H}_2$ as
\begin{equation}
{\cal T}\cdot \omega =\omega^t.
\end{equation}
The group $\Gamma_{\cal T}$ is defined to be the semi-direct product of the modular group $\Gamma$ and $<{\cal T}>$:
\begin{equation}
\Gamma_{\cal T} := \Gamma \, \rtimes <{\cal T}>.
\end{equation}
There is an isomorphism $\Gamma_{\cal T} \cong O^+(L^{2,4})$, and this induces the isomorphism (\ref{H2 isom in 2.2}) \cite{Matsumoto}.
\par Under the isomorphism $\Gamma_{\cal T} \cong O^+(L^{2,4})$, the ring of $O^+(L^{2,4})$-modular forms corresponds to the ring of $\Gamma_{\cal T}$-modular forms of even characteristic \cite{Vinberg2010}, generated by the five modular forms $J_k$ of weights $2k$, $k=2, \cdots, 6$ \cite{CMSE7}. See \cite{CMSE7} for definitions of the modular forms $J_k$. In a special situation, the modular forms $J_2, J_3, J_5, J_6$ restrict to Igusa's generators \cite{Igusa}, $\psi_4, \psi_6, \chi_{10}, \chi_{12}$ (and $J_4$ vanishes in this situation) \cite{CMSE7}.
\par The periods of $H\oplus E_7\oplus E_7$ lattice polarized K3 surfaces $S$ in $H^2(S, \Z)$ determine points in ${\bf H}_2$. The Weierstrass coefficients of such elliptically fibered K3 surfaces were given in terms of $\Gamma_{\cal T}$-modular forms of even characteristic \cite{CMSE7}. 
\par The Weierstrass equation of a K3 surface with $H\oplus E_7\oplus E_7$ lattice polarization is given by \cite{CMSE7}:
\begin{equation}
\label{Weierstrass with param in 2.2}
y^2 = x^3+ (e\, t^4+c\, t^3+a\, t^2) x + t^7+g\, t^6+(d\, e+f)\, t^5+c\, d\, t^4 + b\, t^3.
\end{equation}
Up to some scale factors, the coefficients are given in terms of the modular forms $J_2, J_3, J_4, J_5, J_6$ \cite{CMSE7}:
\begin{eqnarray}
\label{coeffs in 2.2}
c= -J_5(\omega), & d=-\frac{1}{3}J_4(\omega), & e = -3J_2(\omega) \\ \nonumber
f= J_6(\omega), & &  g=-2J_3(\omega) \\ \nonumber 
a=-3 d^2=-\frac{1}{3}J_4(\omega)^2, & & b=-2d^3=\frac{2}{27}J_4(\omega)^3.
\end{eqnarray}
The elliptically fibered K3 surface determines a point in $D_{2,4}$, and this also determines a point in ${\bf H}_2$ under the isomorphism (\ref{H2 isom in 2.2}), which we denote by $\omega$.
\par Now, consider a manifold $M$ and a line bundle $\Lambda$ on $M$, and choose sections $a,b,c,d,e,f, g$ of the line bundles $\Lambda^{\otimes 16}$, $\Lambda^{\otimes 24}$, $\Lambda^{\otimes 10}$, $\Lambda^{\otimes 8}$, $\Lambda^{\otimes 4}$, $\Lambda^{\otimes 12}$ and $\Lambda^{\otimes 6}$, respectively. When the sections $a,b,c,d,e,f,g$ are identified as (\ref{coeffs in 2.2}), because $\Gamma_{\cal T}$ is isomorphic to $O^+(L^{2,4})$, the compactification on $M$ (which is the 2-torus when we consider 8D heterotic strings) gives a heterotic string theory with $O^+(L^{2,4})$-symmetry. The moduli space of eight-dimensional heterotic strings on the 2-torus $T^2$ decomposes into the product of the complex structure moduli, the Wilson line expectation values and K\"ahler moduli, in a suitable limit \cite{NSWheterotic}. The complex structure moduli, the Wilson line expectation values and K\"ahler moduli are mixed under the $O^+(L^{2,4})$-symmetry. This represents the construction of non-geometric heterotic strings with $\mathfrak{e}_7\mathfrak{e}_7$ gauge algebra in \cite{CMSE7}.
\par The locus in the moduli in which the singularity ranks of elliptic K3 surfaces are enhanced satisfies 5-brane solutions on the heterotic side. The generic 5-brane solutions of non-geometric heterotic strings with $\mathfrak{e}_7\mathfrak{e}_7$ gauge algebra are discussed in \cite{CMSE7}.

\vspace{5mm}

\par Elliptic K3 surfaces with the lattice polarization $H\oplus E_7 \oplus E_7$ were described in \cite{ClingherDoran2011} as the minimal resolution of the quartic hypersurfaces in $\P^3$ given by the following equations:
\begin{equation}
\label{hypersurface in 2.2}
Y^2ZW-4X^3Z+3\alpha XZW^2+\beta ZW^3+\gamma XZ^2W- \frac{1}{2} (\zeta W^4+\delta Z^2W^2)+\varepsilon XW^3=0,
\end{equation}
where $[X:Y:Z:W]$ are homogeneous coordinates on $\P^3$. $\alpha, \beta, \gamma, \delta, \varepsilon, \zeta$ are parameters, and $(\gamma, \delta)\ne (0,0)$, and $(\varepsilon, \zeta)\ne (0,0)$.
\par Making the following substitutions 
\begin{eqnarray}
X & = & tx \\ \nonumber
Y & = & y \\ \nonumber
W & = & 4t^3 \\ \nonumber 
Z & = & 4t^4
\end{eqnarray}
yields the Weierstrass equation with two type $III^*$ fibers as follows \cite{CMSE7} :
\begin{equation}
\label{first fibration in 2.2}
y^2 = x^3 + 4t^3\, (\gamma t^2 -3\alpha t+\varepsilon)\, x-8t^5\, (\delta t^2+2\beta t+\zeta).
\end{equation}
Type $III^*$ fibers are at $t=0$ and at $t=\infty$.
\par K3 surface with the lattice polarization $H\oplus E_7 \oplus E_7$ given by (\ref{hypersurface in 2.2}) always admits another fibration with a type $II^*$ fiber and a type $I^*_2$ fiber, as shown in \cite{CMSE7}, and the Weierstrass equation of this fibration is:
\begin{equation}
\label{second fibration in 2.2}
\begin{split}
y^2 = & x^3 - \frac{1}{3} \big[9\alpha t^4+3(\gamma\zeta +\delta\varepsilon)\, t^3+ (\gamma\varepsilon)^2\, t^2 \big] x\\
 & + \frac{1}{27} \big[27t^7 - 54\beta t^6 + 27 (\alpha\gamma\varepsilon+\delta\zeta)\, t^5+9\gamma\varepsilon (\delta\varepsilon +\gamma\zeta)\, t^4+2 (\gamma\varepsilon)^3\, t^3 \big].
 \end{split}
\end{equation}
The Weierstrass equation (\ref{second fibration in 2.2}) was used in \cite{CMSE7} to construct eight-dimensional non-geometric heterotic strings with unbroken $\mathfrak{e}_7\mathfrak{e}_7$. (Compare the equation (\ref{second fibration in 2.2}) with the equation (\ref{Weierstrass with param in 2.2}).) Although the presence of two $E_7$ singularities in the Weierstrass equation is explicit in the equation (\ref{first fibration in 2.2}), as stated in \cite{CMSE7}, the Weierstrass equation (\ref{first fibration in 2.2}) does not necessarily extend over the entire parameter space. For this reason, the Weierstrass equation (\ref{second fibration in 2.2}) was instead used to construct non-geometric heterotic strings with unbroken $\mathfrak{e}_7\mathfrak{e}_7$ in \cite{CMSE7}.

\par The K3 surface with the lattice polarization $H\oplus E_7 \oplus E_7$ (\ref{hypersurface in 2.2}) also admits another elliptic fibration further, the singular fibers of which include a type $I_8^*$ fiber (or worse) \cite{CMSE7}. This alternate fibration relates to $SO(32)$ heterotic string. The Weierstrass equation of this fibration is obtained by making the following substitutions into the equation (\ref{hypersurface in 2.2}) \cite{CMSE7}:
\begin{eqnarray}
X & = & tx^3 \\ \nonumber
Y & = & \sqrt{2}x^2 y \\ \nonumber
W & = & 2x^3 \\ \nonumber 
Z & = & 2x^2 (-\varepsilon t+\zeta).
\end{eqnarray}
The Weierstrass equation is \cite{CMSE7} :
\begin{equation}
y^2 = x^3 + A x^2 + B x,
\end{equation}
where
\begin{eqnarray}
A & = & t^3-3\alpha t-2\beta \\ \nonumber
B & = & (\gamma t-\delta) (\varepsilon t-\zeta).
\end{eqnarray}
The discriminant is given by 
\begin{equation}
\Delta = B^2 \, (A^2 -4B).
\end{equation}

\subsection{Extremal K3 surfaces}
\label{ssec2.3}
By the Shioda--Tate formula \cite{Shiodamodular, Tate1, Tate2}, the following equality holds for an elliptic surface $S$ with a global section:
\begin{equation}
{\rm rk} ADE + {\rm rk} MW +2 = \rho(S).
\end{equation}
We have used rk $ADE$ to denote the rank of the singularity type of an elliptic surface $S$. The Picard number $\rho(S)$ ranges from 2 to 20 for an elliptic K3 surface with a section. Thus, the rank of the singularity of an elliptic K3 surface $S$ with a section is bounded by:
\begin{equation}
{\rm rk} ADE = \rho(S)-2-{\rm rk} MW \le 18-{\rm rk} MW.
\end{equation}
Therefore, the rank of the singularity of an elliptic K3 surface $S$ with a section can be 18 at the highest, and this value is achieved precisely when the Picard number attains the highest value 20, and the Mordell--Weil rank is 0. Physically, this means that the rank of the gauge group on F-theory compactification on an elliptic K3 surface is at most 18, and when the non-Abelian gauge group has the rank 18, it does not have a $U(1)$ gauge field.
\par K3 surfaces with Picard number 20 are called {\it attractive} K3 surfaces\footnote{We refer to complex K3 surfaces with the highest Picard number 20 as attractive K3 surfaces, following the convention for the term used in \cite{M}.}. The complex structure moduli of the attractive K3 surfaces is known to be parametrized by three integers. The transcendental lattice $T(S)$ of a K3 surface $S$ is the orthogonal complement of the N\'eron--Severi lattice $NS(S)$ inside the K3 lattice $\Lambda_{K3}$, and the transcendental lattices $T(S)$ of attractive K3 surfaces are positive-definite, even 2 $\times$ 2 lattices. The complex structure of an attractive K3 surface is determined by the transcendental lattice \cite{PS-S, SI}. The intersection form of the transcendental lattice of an attractive K3 surface can be transformed into the following form under the $GL_2(\Z)$ action:
\begin{equation}
\label{general intersection form in 2.3}
\begin{pmatrix}
2a & b \\
b & 2c \\
\end{pmatrix}.
\end{equation}
Here $a,b,c$ are integers, $a,b,c\in\Z$, and satisfy the relations: 
\begin{equation}
a \ge c \ge b  \ge 0.
\end{equation}
Thus, the triplet of integers, $a,b,c$, parameterizes the complex structure moduli of the attractive K3 surfaces. We denote an attractive K3 surface, whose transcendental lattice has the intersection form (\ref{general intersection form in 2.3}) as $S_{[2a \hspace{1mm} b \hspace{1mm} 2c]}$ in this note.
\par An elliptic attractive K3 surfaces with a section is said to be {\it extremal} when it has Mordell--Weil rank 0. This condition is equivalent to an elliptic K3 surface with a section having singularity rank 18. Thus, the non-Abelian gauge group forming in F-theory compactification on an elliptic K3 surface has rank 18 precisely when the K3 surface is extremal. In Section \ref{sec3}, we study the points in the moduli of eight-dimensional non-geometric heterotic strings with unbroken $\mathfrak{e}_7\mathfrak{e}_7$ algebra at which the non-Abelian gauge groups are enhanced to rank 18 on the F-theory side. Elliptic K3 surfaces on the F-theory side become extremal at these points.
\par Elliptically fibered K3 surfaces generally admit distinct elliptic fibrations\footnote{Genus-one fibered K3 surfaces in general admit both genus-one fibrations without a section, as well as elliptic fibrations with a section. However, as shown in \cite{Keum}, the attractive K3 surface with discriminant four, $S_{[2 \hspace{1mm} 0 \hspace{1mm} 2]}$, only admits elliptic fibrations with a global section. The authors of \cite{KimuraMizoguchi} discussed F-theory compactification on the surface $S_{[2 \hspace{1mm} 0 \hspace{1mm} 2]}$, in relation to the appearances of the $U(1)$ factor.}, and distinct elliptic fibrations have different singularity types and different Mordell--Weil groups. Physically, this means that the gauge groups and $U(1)$ gauge fields that arise in F-theory compactification on an elliptic K3 surface with the fixed complex structure vary, because there still remains freedom to choose a fibration structure among the distinct choices of elliptic fibrations of that elliptic K3 surface\footnote{This point is discussed in \cite{BKW}.}. 
\par The attractive K3 surface whose transcendental lattice has the intersection form
\begin{equation}
\begin{pmatrix}
2 & 0 \\
0 & 2 \\
\end{pmatrix}
\end{equation}
is particularly relevant to the contents of this study. The elliptic fibrations of the attractive K3 surface $S_{[2 \hspace{1mm} 0 \hspace{1mm} 2]}$ were classified in \cite{Nish} and there are 13 types. We list these 13 types of elliptic fibration of the attractive K3 surface $S_{[2 \hspace{1mm} 0 \hspace{1mm} 2]}$ in Appendix \ref{secA}.

\section{Special points in the moduli of eight-dimensional non-geometric heterotic strings and F-theory duals with enhanced gauge groups}
\label{sec3}
\subsection{Summary}
\label{ssec3.1}
There are finitely many points in the moduli of eight-dimensional non-geometric heterotic strings with unbroken $\mathfrak{e}_7\mathfrak{e}_7$ algebra, at which the non-Abelian gauge groups on the F-theory side are enhanced to rank 18. 
\par In Section \ref{ssec3.2}, we show that the heterotic strings at these special points in the moduli can be described as deformations of the stable degenerations, as a result of the coincident 7-branes on the F-theory side. This effect can be seen as the insertion of 5-branes in the heterotic language. We also discuss applications to $SO(32)$ heterotic strings.
\par As stated in Section \ref{ssec2.3}, K3 surfaces become extremal on the F-theory side at these points in the moduli. The complex structures of the extremal K3 surfaces were classified in \cite{SZ}, and using this result, the complex structures of the extremal K3 surfaces at these points in the moduli can be determined. This enables us to determine the Weierstrass equations of the extremal K3 surfaces that appear as compactification spaces on the F-theory side in the moduli. By using this approach, we study the physics of the theories at the enhanced special points in the moduli. 
\par In eight-dimensional $E_8\times E_8$ heterotic strings on the 2-torus $T^2$, only the gauge groups up to $E_8\times E_8\times U(1)^4$ can arise in the perturbative description \cite{Kimura1810}. This implies that the heterotic dual of F-theory on an extremal elliptic K3 surface with the non-Abelian gauge group of rank 18 does not allow for the perturbative interpretation of the gauge group. This can reflect some non-perturbative aspects of the non-geometric heterotic strings. 

\subsection{F-theory on extremal K3 surfaces and non-geometric heterotic duals in the moduli}
\label{ssec3.2}
We discuss the points in the moduli of non-geometric heterotic strings with unbroken $\mathfrak{e}_7\mathfrak{e}_7$ algebra, at which the non-Abelian gauge symmetries on the F-theory side are enhanced to rank 18. K3 surfaces as compactification spaces on the F-theory side become extremal at these points. There are finitely many such points in the moduli, and the complex structures and the singularity types of the extremal K3 surfaces that appear in the moduli can be determined from Table 2 of \cite{SZ}. Among these, those of the singularity types, which include $E_8 E_7$, are studied in \cite{Kimura1810}. We do not discuss these extremal K3 surfaces in this note. Instead, we discuss the extremal K3 surfaces that belong to the moduli, the singularity types of which include $E_7^2$ \footnote{The singularity types of extremal K3 surfaces can also be enhanced to $E_8D_{7}$, as discussed in \cite{CMSE7}.}. 
\par The singularity types of the extremal K3 surfaces in the moduli of K3 surfaces with $H\oplus E_7 \oplus E_7$ lattice polarization, which do not include $E_8$, are as follows \cite{SZ}:
\begin{equation}
E_7^2 A_3 A_1, \hspace{2mm} E_7^2D_4, E_7^2 A_4, \hspace{2mm} E_7^2 A_2^2.
\end{equation}
We study F-theory on the extremal K3 surfaces possessing the first two singularity types in this note. 
\par Because the perturbative eight-dimensional heterotic strings on $T^2$ can have up to $E_8\times E_8\times U(1)^4$ gauge group, the heterotic duals of F-theory on these extremal K3 surfaces do not allow for the perturbative interpretation of these gauge groups \cite{Kimura1810}. As we demonstrate in Sections \ref{sssec3.2.1} and \ref{sssec3.2.2}, these theories can be seen as deformations of the stable degenerations as a result of the coincident 7-branes on the F-theory side. These theories satisfy multiple 5-brane solutions on the heterotic side.

\subsubsection{Extremal K3 surface with $E_7^2 A_3 A_1$ singularity}
\label{sssec3.2.1}
The complex structure of the extremal K3 surface with $E_7^2 A_3 A_1$ singularity is uniquely determined, and its transcendental lattice has the following intersection form \cite{SZ}:
\begin{equation}
\begin{pmatrix}
4 & 0 \\
0 & 2 \\
\end{pmatrix}.
\end{equation}
Therefore, the attractive K3 surface $S_{[4 \hspace{1mm} 0 \hspace{1mm} 2]}$\footnote{The elliptic fibrations and the Weierstrass equations of the attractive K3 surface $S_{[4 \hspace{1mm} 0 \hspace{1mm} 2]}$ were obtained in \cite{BLe}.} admits an extremal fibration with the singularity type $E_7^2 A_3 A_1$, and F-theory on this extremal fibration has non-geometric heterotic dual with unbroken $\mathfrak{e}_7\mathfrak{e}_7$. The Weierstrass form of this extremal fibration can be found in \cite{KRES} as
\begin{equation}
\label{extremal fibration 402 in 3.2.1}
y^2=x^3-\frac{9}{16}(t^2+s^2+\frac{10}{3}ts)\, t^3s^3\, x+\frac{9}{4}t^5s^5\, (\frac{1}{4}t^2+\frac{1}{4}s^2+\frac{7}{18}ts),
\end{equation}
the singular fibers of which consist of two type $III^*$ fibers, a type $I_4$ fiber, and a type $I_2$ fiber. The above Weierstrass equation was obtained in \cite{KRES} as the quadratic base change of an extremal rational elliptic surface. Geometrically, the quadratic base change of a rational elliptic surface is to glue a pair of identical rational elliptic surfaces. Extremal rational elliptic surfaces are the rational elliptic surfaces with a global section, the singularity types of which have rank 8. The types of singular fibers of the extremal rational elliptic surfaces were classified in \cite{MP}. The fiber types of the extremal rational elliptic surfaces are listed in Appendix \ref{secB}.
\par The complex structures of extremal rational elliptic surfaces are uniquely specified by the fiber types, except those with two fibers of type $I^*_0$ (see \cite{MP}). The complex structures of extremal rational elliptic surfaces with two fibers of type $I_0^*$ depend on the $j$-invariants of the fibers. The $j$-invariant $j$ of an extremal rational elliptic surface with two type $I_0^*$ fibers is constant over the base, and the fixed $j$ specifies the complex structure \cite{MP}. In this study, we denote, for example, the extremal rational elliptic surface with a type $III^*$ fiber and a type $III$ fiber as $X_{[III, \hspace{1mm} III^*]}$. We simply use $n$ to denote a singular fiber of type $I_n$ and $m^*$ to represent a fiber of type $I^*_m$. The extremal rational elliptic surface with a type $III^*$ fiber, a type $I_2$ fiber and a type $I_1$ fiber is denote as $X_{[III^*, \hspace{1mm} 2, 1]}$. Because the complex structure of an extremal rational elliptic surface with two type $I_0^*$ fibers depends on the $j$-invariant of the elliptic fibers, we use $X_{[0^*, \hspace{1mm} 0^*]}(j)$ to denote this extremal rational elliptic surface. 
\par Now we demonstrate that F-theory on an extremal K3 surface (\ref{extremal fibration 402 in 3.2.1}) can be seen as a deformation of stable degeneration, owing to an effect of coincident 7-branes. As deduced in \cite{KRES}, the K3 extremal fibration (\ref{extremal fibration 402 in 3.2.1}) is obtained as the quadratic base change of the extremal rational elliptic surface $X_{[III^*, \hspace{1mm} 2, 1]}$ in which two type $I_2$ fibers and two type $I_1$ fibers collide. Whereas the quadratic base change of a rational elliptic surface generally yields an elliptic K3 surface, with twice as many singular fibers as the original rational elliptic surface, at the special limits at which singular fibers collide, the singularity type of the resulting K3 surface is enhanced. As discussed in \cite{KRES}, two identical extremal rational elliptic surfaces $X_{[III^*, \hspace{1mm} 2, 1]}$ are glued together to yield an elliptic K3 surface, which we denote as $S_1$, the singular fibers of which consist of two type $III^*$ fibers, two type $I_2$ fibers, and two type $I_1$ fibers. In the special limit at which 7-branes over which type $I_2$ fiber lies coincide with those over which type $I_2$ fiber lies, and 7-brane over which type $I_1$ fiber lies coincides with 7-brane over which type $I_1$ fiber lies, two type $I_2$ fibers are enhanced to type $I_4$ fiber, and two type $I_1$ fibers are enhanced to type $I_2$ fiber. Because a K3 surface with two type $III^*$ fibers, a type $I_4$ fiber, and a type $I_2$ fiber has the singularity type $E_7^2 A_3 A_1$, the K3 surface $S_1$ deforms and it becomes an extremal K3 surface (\ref{extremal fibration 402 in 3.2.1}) in this limit. In short, F-theory on the extremal K3 surface (\ref{extremal fibration 402 in 3.2.1}) can be seen as deformation of the stable degeneration because of the coincident 7-branes.
\par As the singularity rank of a rational elliptic surface is up to 8, the non-Abelian gauge group that arises on F-theory on a generic K3 surface obtained as the reverse of the stable degeneration has rank up to 16. Here, by generic we mean a situation in which singular fibers of rational elliptic surfaces do not collide when they are glued together to yield an elliptic K3 surface. When the large radius limit is taken, the heterotic dual of this compactification admits a geometric interpretation. In special situations in which singular fibers collide, 7-branes become coincident and the singularity ranks of the resulting K3 surfaces enhance to become greater than 16. The gauge groups of the heterotic duals of F-theory compactifications on these K3 surfaces do not allow for geometric interpretation.
\par The Mordell--Weil group of the K3 extremal fibration (\ref{extremal fibration 402 in 3.2.1}) is isomorphic to $\Z_2$ \cite{SZ, BLe}. Thus, the gauge group that arises in F-theory compactification on this extremal K3 surface is \cite{KRES}
\begin{equation}
E_7 \times E_7 \times SU(4) \times SU(2) / \Z_2.
\end{equation}
\par Comparing the Weierstrass equation (\ref{extremal fibration 402 in 3.2.1}) with equation (\ref{first fibration in 2.2}), we find that the following substitutions:
\begin{eqnarray}
\label{substitution 402 in 3.2.1}
\alpha & = & \frac{5}{32} \\ \nonumber
\beta & = & -\frac{7}{128} \\ \nonumber
\gamma = \varepsilon & = & -\frac{9}{64} \\ \nonumber
\delta = \zeta & = & -\frac{9}{128}   
\end{eqnarray}
into equation (\ref{first fibration in 2.2}) yield the Weierstrass equation (\ref{extremal fibration 402 in 3.2.1}). Plugging the substitutions (\ref{substitution 402 in 3.2.1}) into the equation of the alternate fibration (\ref{second fibration in 2.2}), we obtain the following equation:
\begin{equation}
\label{alternate 402 in 3.2.1}
y^2=x^3-\frac{1}{3}\, (10t^4+8t^3+t^2)\, x+t^7+\frac{56}{27}t^6+\frac{26}{9}t^5+\frac{8}{9}t^4+\frac{2}{27}t^3,
\end{equation}
with the discriminant
\begin{equation}
\label{disc alternate 402 in 3.2.1}
\Delta \sim t^{11} (t+2)^2 (27t+4).
\end{equation}
From equations (\ref{alternate 402 in 3.2.1}) and (\ref{disc alternate 402 in 3.2.1}), we find that the fibration (\ref{alternate 402 in 3.2.1}) has a type $II^*$ fiber at $t=\infty$, a type $I_5^*$ fiber at $t=0$, a type $I_2$ fiber at $t=-2$, and a type $I_1$ fiber at $t=-4/27$. Thus, the fibration (\ref{alternate 402 in 3.2.1}) has singularity type $E_8 D_9 A_1$, and because the singularity type has rank 18, we deduce that this fibration is also extremal. Therefore, we find that the attractive K3 surface $S_{[4 \hspace{1mm} 0 \hspace{1mm} 2]}$ also admits an extremal fibration (\ref{alternate 402 in 3.2.1}) with singularity type $E_8 D_9 A_1$. This agrees with the results in \cite{SZ, BLe}.  

\subsubsection{Extremal K3 surface with $E_7^2 D_4$ singularity}
\label{sssec3.2.2}
The complex structure of the extremal K3 surface with the singularity type $E_7^2 D_4$ is uniquely determined, and the intersection form of the transcendental lattice is \cite{SZ}:
\begin{equation}
\begin{pmatrix}
2 & 0 \\
0 & 2 \\
\end{pmatrix}.
\end{equation}
The Weierstrass equation of this extremal fibration of the attractive K3 surface $S_{[2 \hspace{1mm} 0 \hspace{1mm} 2]}$ is given as follows:
\begin{equation}
\label{extremal 202 in 3.2.2}
y^2=x^3+ 4t^3(t-s)^2 \, s^3 \, x,
\end{equation}
with the discriminant 
\begin{equation}
\Delta \sim t^9s^9 \, (t-s)^6.
\end{equation}
$[t:s]$ in the equation (\ref{extremal 202 in 3.2.2}) denotes the homogeneous coordinate of the base $\P^1$. Two type $III^*$ are at $[t:s]=[0:1]$ and $[1:0]$, and a type $I^*_0$ fiber is at $[t:s]=[1:1]$. 
\par As shown in \cite{KRES}, the extremal K3 fibration (\ref{extremal 202 in 3.2.2}) can be seen as deformation of the stable degeneration. Gluing two identical extremal rational elliptic surfaces $X_{[III, \hspace{1mm} III^*]}$ yields an elliptic K3 surface, $S_2$, the singular fibers of which have two type $III^*$ fibers and two type $III$ fibers. This is technically given by a generic quadratic base change of the extremal rational elliptic surface $X_{[III, \hspace{1mm} III^*]}$, and this is the reverse of the stable degeneration. In a special limit at which two type $III$ fibers collide, the elliptic K3 surface $S_2$ deforms to yield the extremal fibration (\ref{extremal 202 in 3.2.2}) of the attractive K3 surface $S_{[2 \hspace{1mm} 0 \hspace{1mm} 2]}$ \cite{KRES}. 7-branes over which type $III$ fiber lies coincide with those over which type $III$ fiber lies in this limit, at which colliding two type $III$ fibers are enhanced to a type $I_0^*$ fiber. Therefore, F-theory on the extremal K3 surface (\ref{extremal 202 in 3.2.2}) can be seen as deformation of the stable degeneration as the consequence of the coincident 7-branes.
\par The Mordell--Weil group of the extremal elliptic fibration (\ref{extremal 202 in 3.2.2}) is isomorphic to $\Z_2$ \cite{Nish, SZ}. Therefore, the gauge group on F-theory compactification on the extremal fibration (\ref{extremal 202 in 3.2.2}) is \cite{KRES}
\begin{equation}
E_7 \times E_7 \times SO(8) / \Z_2.
\end{equation}
\par Comparing the equation (\ref{extremal 202 in 3.2.2}) with the equation (\ref{first fibration in 2.2}), we find that the following substitutions 
\begin{eqnarray}
\label{substitution 202 in 3.2.2}
\alpha & = & \frac{2}{3} \\ \nonumber
\beta = \delta = \zeta & = & 0 \\ \nonumber
\gamma = \varepsilon & = & 1  
\end{eqnarray}
into (\ref{first fibration in 2.2}) yield the Weierstrass equation (\ref{extremal 202 in 3.2.2}). 
\par By plugging the substitutions (\ref{substitution 202 in 3.2.2}) into the equation (\ref{second fibration in 2.2}), we obtain the following Weierstrass equation:
\begin{equation}
\label{alternate 202 in 3.2.2}
y^2=x^3-\frac{1}{3}t^2(6t^2+1)x+\frac{1}{27}t^3(27t^4+18t^2+2),
\end{equation}
with the discriminant 
\begin{equation}
\label{disc alternate 202 in 3.2.2}
\Delta \sim t^{12} (27t^2+4).
\end{equation}
We can confirm from the equations (\ref{alternate 202 in 3.2.2}) and (\ref{disc alternate 202 in 3.2.2}) that this alternate fibration in fact has a type $II^*$ fiber at $t=\infty$, a type $I_6^*$ fiber at $t=0$, and two type $I_1$ fibers at the roots of $27t^2+4=0$. Thus, the singularity type of the alternate fibration is $E_8 D_{10}$, and we find that this fibration is also extremal. This gives the Weierstrass equation of the fibration no. 2 in Table \ref{tablefibrations202 in A} of the attractive K3 surface $S_{[2 \hspace{1mm} 0 \hspace{1mm} 2]}$ in Appendix \ref{secA}. 

\vspace{5mm}

\par Double cover of $\P^1\times \P^1$ ramified along a bidegree (4,4) curve, given by the following equation:
\begin{equation}
\label{genusone K3 in 3.2.2}
\tau^2=(t-\alpha_1)^3(t-\alpha_2)\, x^4+(t-\alpha_3)^3 (t-\alpha_2)
\end{equation}
yields a genus-one fibered K3 surface lacking a global section, but admitting a bisection, and this K3 surface was considered in \cite{K2} in the context of F-theory compactifications on genus-one fibrations without a global section. $x$ denotes the inhomogeneous coordinate of the first $\P^1$, and $t$ denotes the inhomogeneous coordinate of the second $\P^1$, in the product $\P^1\times \P^1$, respectively. $\alpha_1$, $\alpha_2$, $\alpha_3$ are distinct points in $\P^1$. $\alpha$'s are superfluous parameters, and these can be mapped to:
\begin{equation}
\alpha_1=0, \hspace{2mm} \alpha_2=1, \hspace{2mm} \alpha_3=\infty
\end{equation}
under some appropriate automorphism of the base $\P^1$. The K3 genus-one fibration (\ref{genusone K3 in 3.2.2}) has two type $III^*$ fibers at $t=\alpha_1, \alpha_3$, and a type $I_0^*$ fiber at $t=\alpha_2$ \cite{K2}. 
\par The Jacobian fibration of the K3 genus-one fibration (\ref{genusone K3 in 3.2.2}) gives the extremal K3 elliptic fibration (\ref{extremal 202 in 3.2.2}), as demonstrated in \cite{K2}. Utilizing this fact, in Section \ref{sec4} we build an elliptically fibered Calabi--Yau 3-fold, by fibering the K3 genus-one fibration (\ref{genusone K3 in 3.2.2}) over the base $\P^1$, then taking the Jacobian fibration of it. It turns out that the resulting Calabi--Yau 3-fold is a fibration of the extremal K3 surface (\ref{extremal 202 in 3.2.2}) over the base $\P^1$. We also build a family of elliptic Calabi--Yau 3-folds which includes this Calabi--Yau 3-fold. These constructions will be discussed in detail in Section \ref{sec4}.

\subsection{Applications to $SO(32)$ heterotic strings}
\label{ssec3.3}
As reviewed in Section \ref{ssec2.2}, the K3 surface (\ref{hypersurface in 2.2}) with $H\oplus E_7 \oplus E_7$ lattice polarization admits another elliptic fibration, the singular fibers of which include a type $I_8^*$ fiber, given by the Weierstrass equation:
\begin{equation}
\label{third fibration in 3.3}
y^2=x^3+ (t^3-3\alpha t-2\beta)\, x^2+(\gamma t-\delta)(\varepsilon t-\zeta)\, x
\end{equation}
with the discriminant 
\begin{equation}
\Delta \sim (\gamma t-\delta)^2(\varepsilon t-\zeta)^2 \big[(t^3-3\alpha t-2\beta)^2-4(\gamma t-\delta)(\varepsilon t-\zeta)\big].
\end{equation}
Therefore, there is a birational map that transforms the elliptic fibration with two type $III^*$ fibers (\ref{first fibration in 2.2}) into an alternate fibration (\ref{third fibration in 3.3}) with a type $I_8^*$ fiber (or worse). Using this birational map, we send the extremal K3 elliptic fibrations with two $E_7$ singularities studied in Section \ref{ssec3.2} to another fibration with a type $I_8^*$ fiber (or worse). This relates to $SO(32)$ heterotic strings. 
\vspace{5mm}
\par As we saw previously in Section \ref{sssec3.2.2}, the Weierstrass equation of the extremal fibration of the attractive K3 surface $S_{[2 \hspace{1mm} 0 \hspace{1mm} 2]}$ with singularity type $E_7^2 D_4$ is given by (\ref{extremal 202 in 3.2.2}), with
\begin{eqnarray}
\label{substitution 202 in 3.3}
\alpha & = & \frac{2}{3} \\ \nonumber
\beta = \delta = \zeta & = & 0 \\ \nonumber
\gamma = \varepsilon & = & 1.  
\end{eqnarray}
By plugging these values (\ref{substitution 202 in 3.3}) into the alternate fibration (\ref{third fibration in 3.3}), we obtain the Weierstrass equation as
\begin{equation}
\label{transf 202 in 3.3}
y^2=x^3+t(t^2-2)\, x^2+t^2\, x,
\end{equation}
with the discriminant 
\begin{eqnarray}
\Delta & \sim & t^4 \, \big( t^2(t^2-2)^2-4t^2 \big) \\ \nonumber
& = & t^8\, (t^2-4).
\end{eqnarray}
In the homogeneous form, the discriminant is
\begin{equation}
\label{disc tranf 202 in 3.3}
\Delta \sim t^8 s^{14}\, (t^2-4s^2).
\end{equation}
From equations (\ref{transf 202 in 3.3}) and (\ref{disc tranf 202 in 3.3}), we deduce that the alternate fibration (\ref{transf 202 in 3.3}) has a type $I_8^*$ fiber at $[t:s]=[1:0]$, a type $I_2^*$ fiber at $[t:s]=[0:1]$, and two type $I_1$ fibers at $[t:s]=[2:1], [-2:1]$. Thus, the alternate fibration (\ref{transf 202 in 3.3}) has singularity type $D_{12}D_6$, and we find that this fibration is also extremal. We conclude that the alternate fibration (\ref{transf 202 in 3.3}) yields fibration no.\ 8 in Table \ref{tablefibrations202 in A} in Appendix \ref{secA} of the attractive K3 surface $S_{[2 \hspace{1mm} 0 \hspace{1mm} 2]}$.
\par The Mordell--Weil group of the fibration (\ref{transf 202 in 3.3}) is isomorphic to $\Z_2$ (see \cite{Nish, SZ}); therefore, the gauge group on F-theory compactification on the fibration (\ref{transf 202 in 3.3}) is
\begin{equation}
SO(24)\times SO(12)/ \Z_2.
\end{equation}

\vspace{5mm}

\par The attractive K3 surface $S_{[2 \hspace{1mm} 0 \hspace{1mm} 2]}$ has another extremal fibration with the singularity type $D_8^2 A_1^2$ \cite{Nish}. (This is fibration no.9 in Table \ref{tablefibrations202 in A} in Appendix \ref{secA}.) As deduced in \cite{KRES}, the Weierstrass equation of this extremal fibration is given as follows:
\begin{equation}
\label{D8 202 in 3.3}
y^2=x^3 -3t^2s^2\, (t^4+s^4-t^2s^2)\, x + (t^2+s^2)\, t^3s^3\, (2t^4-5t^2s^2+2s^4),
\end{equation}
with the discriminant 
\begin{equation}
\Delta \sim t^{10}s^{10}\, (t-s)^2\, (t+s)^2.
\end{equation}
Type $I^*_4$ fibers are at $[t:s]=[1:0], [0:1]$, and type $I_2$ fibers are at $[t:s]=[1:1], [1:-1]$. 
\par As shown in \cite{KRES}, extremal fibration (\ref{D8 202 in 3.3}) can be seen as deformation of the stable degeneration in which two extremal rational elliptic surfaces $X_{[4^*, \hspace{1mm} 1,1]}$ are glued together. Gluing of two extremal rational elliptic surfaces $X_{[4^*, \hspace{1mm} 1,1]}$ yields an elliptic K3 surface, which we denote by $S_3$, the singular fibers of which are two type $I_4^*$ fibers and four type $I_1$ fibers. In a limit at which two pairs of type $I_1$ fibers collide, type $I_1$ fibers collide and they are enhanced to a type $I_2$ fiber. In this limit, K3 surface $S_3$ deforms to yield the attractive K3 surface with the extremal fibration (\ref{D8 202 in 3.3}) \cite{KRES}. Therefore, extremal fibration (\ref{D8 202 in 3.3}) can be seen as deformation of the stable degeneration, as a result of coincident 7-branes over which type $I_1$ fibers lie.
\par The Mordell--Weil group of the fibration (\ref{D8 202 in 3.3}) is isomorphic to $\Z_2\times\Z_2$ \cite{Nish, SZ}; therefore, the gauge group on F-theory compactification on the fibration (\ref{D8 202 in 3.3}) is \cite{KRES}
\begin{equation}
SO(16)\times SO(16)\times SU(2)^2/ \Z_2\times \Z_2.
\end{equation}

\vspace{5mm}
\par We saw previously in Section \ref{sssec3.2.1} that the attractive K3 surface $S_{[4 \hspace{1mm} 0 \hspace{1mm} 2]}$ admits an extremal fibration with the singularity type $E_7^2 A_3 A_1$, and the Weierstrass equation of this fibration is given by (\ref{extremal fibration 402 in 3.2.1}), with
\begin{eqnarray}
\label{substitution 402 in 3.3}
\alpha & = & \frac{5}{32} \\ \nonumber
\beta & = & -\frac{7}{128} \\ \nonumber
\gamma = \varepsilon & = & -\frac{9}{64} \\ \nonumber
\delta = \zeta & = & -\frac{9}{128}.   
\end{eqnarray}
By plugging these into the equation (\ref{third fibration in 3.3}), we obtain an alternate fibration given by:
\begin{equation}
\label{transf 402 in 3.3}
y^2=x^3+(t^3-\frac{15}{32}t+\frac{7}{64})\, x^2+(\frac{9}{64})^2(t-\frac{1}{2})^2 \, x,
\end{equation}
with the discriminant
\begin{equation}
\label{disc transf 402 in 3.3}
\Delta \sim (t-\frac{1}{2})^7 (4t+1)^2 (t+1).
\end{equation}
From the equations (\ref{transf 402 in 3.3}) and (\ref{disc transf 402 in 3.3}), we find that the alternate fibration has a type $I_8^*$ fiber at $t=\infty$, a type $I_1^*$ fiber at $t=\frac{1}{2}$, a type $I_2$ fiber at $t=-\frac{1}{4}$, and a type $I_1$ fiber at $t=-1$. Thus, the alternate fibration (\ref{transf 402 in 3.3}) has the singularity type $D_{12} D_5 A_1$, and this is also extremal. This result agrees with the elliptic fibrations with a section of the attractive K3 surface $S_{[4 \hspace{1mm} 0 \hspace{1mm} 2]}$ obtained in \cite{BLe}. 
\par The Mordell--Weil group of the alternate fibration (\ref{transf 402 in 3.3}) is isomorphic to $\Z_2$ \cite{SZ, BLe}. Therefore, the gauge group on F-theory compactification on the fibration (\ref{transf 402 in 3.3}) is
\begin{equation}
SO(24) \times SO(10) \times SU(2) / \Z_2.
\end{equation}

\section{Jacobian Calabi--Yau 3-folds and F-theory compactifications}
\label{sec4}
\par In this section, we fiber elliptic K3 surfaces over $\P^1$ to yield elliptically fibered Calabi--Yau 3-folds\footnote{In \cite{Nak, DG, G} the elliptic fibrations of 3-folds were discussed.} with a global section, and we study six-dimensional F-theory compactifications with $N=1$ supersymmetry on the constructed Calabi--Yau 3-folds. K3 fibers in this construction include an elliptic K3 surface that belongs to the F-theory side of the moduli of eight-dimensional non-geometric heterotic strings with unbroken $\mathfrak{e}_7\mathfrak{e}_7$ algebra, discussed in Section \ref{sssec3.2.2}. 
\par To be clear, we first consider genus-one fibered K3 surfaces lacking a global section, built as double covers of $\P^1\times \P^1$ ramified along a (4,4) curve. The built genus-one fibered K3 surfaces do not have a global section, but they have a bisection \cite{K2}. We consider higher-dimensional analogs of these K3 surfaces to yield genus-one fibered Calabi--Yau 3-folds, built as double covers of $\P^1\times\P^1\times\P^1$ ramified along a tridegree (4,4,4) surface. As we show in Section \ref{ssec4.1}, the constructed Calabi--Yau 3-folds are genus-one fibered, but they lack a global section. These Calabi--Yau 3-folds still have a bisection \cite{MTsection}. The pullback of $\cal{O}$(1) class in $\P^1$ yields a bisection \cite{Kdisc}. 
\par Taking the Jacobian fibrations of these Calabi--Yau genus-one fibrations yields elliptically fibered Calabi--Yau 3-folds with a global section. The resulting elliptic Calabi--Yau 3-folds are also K3 fibered, and when we tune the coefficients of the defining equation, we obtain Calabi--Yau 3-folds, K3 fibers of which are the attractive K3 surface $S_{[2 \hspace{1mm} 0 \hspace{1mm} 2]}$ that belongs to the F-theory side of the moduli of non-geometric heterotic strings with unbroken $\mathfrak{e}_7\mathfrak{e}_7$ algebra. 
\par We deduce the non-Abelian gauge groups on F-theory compactifications on the Jacobian Calabi--Yau 3-folds. We also perform a consistency check of the obtained gauge groups, by considering the symmetry that the elliptic fibers possess. Highly enhanced gauge groups arise when we choose specific coefficients of the defining equations of the Jacobian Calabi--Yau 3-folds. We determine the Mordell--Weil groups of some specific Calabi--Yau 3-folds, and we deduce the global structures of the gauge groups for F-theory on these spaces. We obtain some models without a $U(1)$ gauge field. Furthermore, we deduce viable candidate matter spectra on F-theory on the constructed elliptically fibered Calabi--Yau 3-folds that satisfy the six-dimensional anomaly cancellation condition.

\subsection{Calabi--Yau 3-folds as double covers, Jacobian fibrations, and the discriminant locus}
\label{ssec4.1}
\par Double covers of the product of projective lines, $\P^1\times \P^1\times \P^1$, ramified over a $(4,4,4)$ surface have the trivial canonical bundles, $K=0$; therefore, they describe Calabi--Yau 3-folds. Fiber of projection onto $\P^1\times\P^1$ is a double cover of $\P^1$ branched over four points, namely, it is an elliptic curve. Thus, projection onto $\P^1\times\P^1$ gives a genus-one fibration. Fiber of projection onto $\P^1$ is a double cover of $\P^1\times\P^1$ ramified along a $(4,4)$ curve, which yields a genus-one fibered K3 surface. Therefore, projection onto $\P^1$ yields a K3 fibration. These K3 surfaces do not have a section, but they have a bisection \cite{K2}.  
\par In this note, we consider in particular the double covers of $\P^1\times\P^1\times\P^1$ given by the following equations:
\begin{equation}
\label{doublecover 3-fold in 4.1}
\tau^2= f_1(t) g_1(u) \, x^4+ f_2(t) g_2(u),
\end{equation}
where $x$ is the inhomogeneous coordinate on the first $\P^1$, and $t$ and $u$ are the inhomogeneous coordinates on the second and the third $\P^1$ in the product $\P^1\times\P^1\times\P^1$. Here $f_1, f_2$ are polynomials in the variable $t$ of degree four, and $g_1, g_2$ are polynomials of degree four in the variable $u$. By splitting the polynomials $f_1, f_2, g_1, g_2$ into linear factors, equation (\ref{doublecover 3-fold in 4.1}) can be rewritten as follows:
\begin{equation}
\label{doublecover rewritten in 4.1}
\tau^2= \Pi_{i=1}^4 (t-\alpha_i)\cdot \Pi_{j=1}^4 (u-\beta_j) \cdot x^4+ \Pi_{k=5}^8 (t-\alpha_k) \cdot \Pi_{l=5}^8 (u-\beta_l).
\end{equation}
K3 fiber of this genus-one fibered Calabi--Yau 3-fold is given by
\begin{equation}
\label{genus-one K3 fiber in 4.1}
\tau^2= \Pi_{i=1}^4 (t-\alpha_i)\cdot x^4+ \Pi_{k=5}^8 (t-\alpha_k).
\end{equation}
As shown in \cite{K2}, K3 fiber (\ref{genus-one K3 fiber in 4.1}) is genus-one fibered, but it does not have a global section. K3 fiber (\ref{genus-one K3 fiber in 4.1}) has a bisection \cite{K2}.
\par Using an argument similar to that in \cite{KCY4}, we can show that the Calabi--Yau 3-fold (\ref{doublecover rewritten in 4.1}) indeed lacks a rational section. Suppose it admits a rational section. Then, it restricts to a K3 fiber, and this gives a global section to the K3 fiber, leading to a contradiction. By an argument similar to those in \cite{K2, KCY4, Kdisc}, the genus-one fibered Calabi--Yau 3-fold (\ref{doublecover rewritten in 4.1}) has a bisection\footnote{Thus, a discrete $\Z_2$ symmetry \cite{MTsection} arises in six-dimensional F-theory compactifications on the genus-one fibered Calabi--Yau 3-folds (\ref{doublecover rewritten in 4.1}).}.
\par We can consider a special situation in which
\begin{equation}
\label{coeff in 4.1}
\alpha_1=\alpha_2=\alpha_3, \hspace{5mm} \alpha_4=\alpha_8, \hspace{5mm} \alpha_5=\alpha_6=\alpha_7.
\end{equation}
This yields a genus-one fibered Calabi--Yau 3-fold
\begin{equation}
\tau^2= (t-\alpha_1)^3(t-\alpha_4)\cdot \Pi_{j=1}^4 (u-\beta_j) \cdot x^4+ (t-\alpha_5)^3(t-\alpha_4) \cdot \Pi_{l=5}^8 (u-\beta_l),
\end{equation}
and K3 fiber given by
\begin{equation}
\label{K3 fiber special in 4.1}
\tau^2= (t-\alpha_1)^3(t-\alpha_4)\, x^4+ (t-\alpha_5)^3(t-\alpha_4).
\end{equation}
This is the genus-one fibered K3 surface (\ref{genusone K3 in 3.2.2}) lacking a section, which we discussed in Section \ref{sssec3.2.2}\footnote{As we stated previously in Section \ref{sssec3.2.2}, $\alpha_1, \alpha_4, \alpha_5$ in equation (\ref{K3 fiber special in 4.1}) are superfluous parameters, and they can be mapped to $0, 1, \infty$ under certain automorphism of the base $\P^1$.}. 
\par The Jacobian fibrations of the genus-one fibered Calabi--Yau 3-folds (\ref{doublecover rewritten in 4.1}) yield elliptically fibered Calabi--Yau 3-folds with a section. The Jacobian fibrations are given by \cite{Muk}
\begin{equation}
\label{Jacobian 3-fold in 4.1}
\tau^2=\frac{1}{4}x^3-\Pi_{i=1}^8 (t-\alpha_i)\cdot \Pi_{j=1}^8 (u-\beta_j)\cdot x.
\end{equation}
Projection onto $\P^1\times\P^1$ gives an elliptic fibration, and projection onto $\P^1$ yields a K3 fibration. K3 fiber of the projection onto $\P^1$ is given by
\begin{equation}
\label{Jacobian K3 fiber in 4.1}
\tau^2=\frac{1}{4}x^3-\Pi_{i=1}^8 (t-\alpha_i)\cdot x.
\end{equation}
\par When parameters $\alpha$ are tuned as in (\ref{coeff in 4.1}), the Weierstrass equation of the Jacobian fibration becomes
\begin{equation}
\label{enhanced Jacobian 3-fold in 4.1}
\tau^2=\frac{1}{4}x^3-(t-\alpha_1)^3(t-\alpha_5)^3(t-\alpha_4)^2\cdot \Pi_{j=1}^8 (u-\beta_j)\, x,
\end{equation}
and K3 fiber (\ref{Jacobian K3 fiber in 4.1}) becomes
\begin{equation}
\label{enhanced Jacobian K3 fiber in 4.1}
\tau^2=\frac{1}{4}x^3-(t-\alpha_1)^3(t-\alpha_5)^3(t-\alpha_4)^2\, x.
\end{equation}
This is the Jacobian fibration of the K3 fiber (\ref{K3 fiber special in 4.1}), and this is the extremal fibration (\ref{extremal 202 in 3.2.2}) of the attractive K3 surface $S_{[2 \hspace{1mm} 0 \hspace{1mm} 2]}$\footnote{Using a coordinate transformation, equation (\ref{enhanced Jacobian K3 fiber in 4.1}) can be replaced with 
$$
\tau^2=x^3+4(t-\alpha_1)^3(t-\alpha_4)^2(t-\alpha_5)^3\, x.
$$
As stated previously, we can send $\alpha_1, \alpha_4, \alpha_5$ to $0,1,\infty$, and this yields (\ref{extremal 202 in 3.2.2}).}, which belongs to the F-theory side of the moduli of the non-geometric heterotic strings with unbroken $\mathfrak{e}_7\mathfrak{e}_7$ algebra. 
\par The obtained Jacobian Calabi--Yau 3-folds (\ref{Jacobian 3-fold in 4.1}) yield fibrations of K3 surfaces (\ref{Jacobian K3 fiber in 4.1}) over $\P^1$, and this family includes fibrations of the extremal K3 surface $S_{[2 \hspace{1mm} 0 \hspace{1mm} 2]}$, which we discussed in Section \ref{sssec3.2.2}, over $\P^1$. 
\par The discriminant of the Calabi--Yau Jacobian fibration (\ref{Jacobian 3-fold in 4.1}) is given by the following equation:
\begin{equation}
\label{disc in 4.1}
\Delta \sim \Pi_{i=1}^8 (t-\alpha_i)^3\cdot \Pi_{j=1}^8 (u-\beta_j)^3.
\end{equation}
The discriminant locus of the Jacobian Calabi--Yau 3-fold (\ref{Jacobian 3-fold in 4.1}) is given by the vanishing of the discriminant (\ref{disc in 4.1}). Genus-one fibered Calabi--Yau 3-fold (\ref{doublecover rewritten in 4.1}) and the Jacobian fibration (\ref{Jacobian 3-fold in 4.1}) have the identical discriminant loci, and the identical configurations of the singular fibers. 
\par From the equation (\ref{disc in 4.1}), we find that the discriminant components of the Jacobian Calabi--Yau 3-fold (\ref{Jacobian 3-fold in 4.1}) are given as follows:
\begin{eqnarray}
A_i & = & \{t=\alpha_i\}  \hspace{5mm} (i=1, \cdots, 8) \\ \nonumber
B_j & = & \{u=\beta_j\}  \hspace{5mm} (j=1, \cdots, 8).
\end{eqnarray}
In F-theory on the Jacobian Calabi--Yau 3-fold (\ref{Jacobian 3-fold in 4.1}), 7-branes are wrapped on these components. Components $A_i$ are isomorphic to $\{{\rm pt}\} \times \P^1$, and components $B_j$ are isomorphic to $\P^1\times \{{\rm pt}\}$. Therefore, these are isomorphic to $\P^1$. The types of the singular fibers and the non-Abelian gauge groups on the 7-branes will be discusses in Section \ref{ssec4.2}.

\subsection{Non-Abelian gauge groups}
\label{ssec4.2}
\par We determine the non-Abelian gauge groups on F-theory on the Jacobian Calabi--Yau 3-folds constructed in Section \ref{ssec4.1}. We also check the consistency of the obtained gauge groups.
\subsubsection{Singular fibers of the Jacobian Calabi--Yau 3-folds, and non-Abelian gauge groups}
\label{sssec4.2.1}
From the Weierstrass equation (\ref{Jacobian 3-fold in 4.1}) of the Jacobian Calabi--Yau 3-fold and the discriminant (\ref{disc in 4.1}), we find that for a generic situation in which the coefficients $\alpha$'s and $\beta$'s are mutually distinct, the types of the singular fibers on the components $A_i$, $i=1, \cdots, 8$, and $B_j$, $j=1, \cdots, 8$, are $III$. In this case, the non-Abelian gauge group that arises on F-theory compactification on the Jacobian Calabi--Yau 3-fold (\ref{Jacobian 3-fold in 4.1}) is
\begin{equation}
SU(2)^{16}.
\end{equation}
\par When two of the coefficients, $\alpha_i$ and $\alpha_k$, become coincident, a pair of type $III$ fibers on the components $A_i$ and $A_k$ collides, and it is enhanced to a type $I^*_0$ fiber. Because the polynomial 
\begin{equation}
x^3 - \Pi_{j=1}^8 (u-\beta_j)\cdot x
\end{equation}
splits into the linear factor and the quadratic factor as:
\begin{equation}
x\, \big(x^2 - \Pi_{j=1}^8 (u-\beta_j) \big),
\end{equation}
type $I^*_0$ fiber is semi-split \cite{BIKMSV}. The non-Abelian gauge group that arises on the 7-branes wrapped on the component $A_i$ is thus enhanced to $SO(7)$ in this situation \footnote{In a special situation in which there are four pairs of identical $\beta$'s, e.g. $\beta_1=\beta_5$, $\beta_2=\beta_6$, $\beta_3=\beta_7$, $\beta_4=\beta_8$, the polynomial splits into three linear factors: 
\begin{equation}
x\, \big(x-\Pi_{j=1}^4 (u-\beta_j) \big) \, \big(x+\Pi_{j=1}^4 (u-\beta_j). \big)
\end{equation}
In this special situation, type $I_0^*$ fiber over the component $A_i$ becomes split, and the gauge group that forms on the 7-branes wrapped on the component $A_i$ is enhanced to $SO(8)$.}
. 
\par When three of the coefficients, $\alpha_i$, $\alpha_k$ and $\alpha_l$, become coincident, type $III$ fibers on the components $A_i$, $A_k$, $A_l$ collide, and they are enhanced to a type $III^*$ fiber. Further enhancement breaks the Calabi--Yau condition, as stated in \cite{K2, KCY4}. An argument similar to that stated previously equally applies to $\beta$'s and the components $B_j$. The results are presented in Table \ref{tablelistgaugegroup in 4.2.1} below.

\begingroup
\renewcommand{\arraystretch}{1.5}
\begin{table}[htb]
\begin{center}
  \begin{tabular}{|c|c|c|} \hline
Component & Fiber type & non-Abel. Gauge Group \\ \hline
$A_{1,\cdots, 8}$ & $III$ & $SU(2)$ \\ \hline
$A_{1,\cdots, 8}$ & $I^*_0$ & $SO(7)$ \\ \hline
$A_{1,\cdots, 8}$ & $III^*$ & $E_7$ \\ \hline 
$B_{1,\cdots, 8}$ & $III$ & $SU(2)$ \\ \hline 
$B_{1,\cdots, 8}$ & $I^*_0$ & $SO(7)$ \\ \hline 
$B_{1,\cdots, 8}$ & $III^*$ & $E_7$ \\ \hline
\end{tabular}
\caption{\label{tablelistgaugegroup in 4.2.1}Fiber types and the gauge groups on the discriminant components.} 
\end{center}
\end{table}  
\endgroup

\par As discussed in Section \ref{ssec4.1}, K3 fiber becomes most enhanced when parameters $\alpha$ are tuned as:
\begin{equation}
\alpha_1=\alpha_2=\alpha_3, \hspace{5mm} \alpha_4=\alpha_8, \hspace{5mm} \alpha_5=\alpha_6=\alpha_7.
\end{equation}
In this case, the non-Abelian gauge group that arises on F-theory compactification on the Jacobian Calabi--Yau 3-fold (\ref{enhanced Jacobian 3-fold in 4.1}) is
\begin{equation}
E_7^2 \times SO(7) \times SU(2)^8.
\end{equation}
K3 fiber becomes the attractive K3 surface $S_{[2 \hspace{1mm} 0 \hspace{1mm} 2]}$ with the singularity type $E_7^2 D_4$ (\ref{enhanced Jacobian K3 fiber in 4.1}) in this situation, and this attractive K3 surface was discussed in Section \ref{sssec3.2.2}. The singularity type of the Jacobian Calabi--Yau 3-fold (\ref{Jacobian 3-fold in 4.1}) is most enhanced, when the following equalities hold among coefficients $\beta$'s further:
\begin{equation}
\beta_1=\beta_2=\beta_3, \hspace{5mm} \beta_4=\beta_8, \hspace{5mm} \beta_5=\beta_6=\beta_7.
\end{equation}
In this case, the Weierstrass equation of the Jacobian Calabi--Yau 3-fold becomes
\begin{equation}
\label{jacobian 3-fold most enhanced in 4.2.1}
\tau^2=\frac{1}{4}x^3-(t-\alpha_1)^3(t-\alpha_5)^3(t-\alpha_4)^2\cdot (u-\beta_1)^3(u-\beta_5)^3(u-\beta_4)^2 \, x.
\end{equation}
The types of the singular fibers over the components, $A_1$, $A_5$, $B_1$, $B_5$, are enhanced to type $III^*$, and the types of the singular fibers over the components $A_4$ and $B_4$ are enhanced to type $I_0^*$. In this situation, the non-Abelian gauge group on the F-theory compactification of the Jacobian Calabi--Yau 3-fold (\ref{jacobian 3-fold most enhanced in 4.2.1}) is enhanced to:
\begin{equation}
E_7^4 \times SO(7)^2.
\end{equation}

\subsubsection{Consistency check of the gauge groups}
\label{sssec4.2.2}
By an argument similar to those given in \cite{K2, KCY4}, smooth genus-one fibers of the Calabi--Yau double covers (\ref{doublecover 3-fold in 4.1}) are invariant under the following transformation:
\begin{equation}
x  \rightarrow  e^{\frac{2 \pi i}{4}}\cdot x.
\end{equation}
We find from this that genus-one fibers of the Calabi--Yau double covers (\ref{doublecover 3-fold in 4.1}) possess complex multiplication of order 4; therefore, the generic genus-one fiber of the Calabi--Yau double cover (\ref{doublecover 3-fold in 4.1}) has j-invariant 1728. This requires the singular fibers to have j-invariant 1728 \cite{K2, KCY4}. Because the types of the singular fibers of Calabi--Yau genus-one fibration and those of the Jacobian fibration are identical, this means that the singular fibers of the Jacobian Calabi--Yau 3-fold (\ref{Jacobian 3-fold in 4.1}) also have j-invariant 1728. 
\par According to Table \ref{tablemonodromy in 2.1} in Section \ref{ssec2.1}, the types of the singular fibers with j-invariant 1728 are: $III$, $III^*$, and $I^*_0$. (j-invariant of type $I^*_0$ fiber can take the value 1728.) Thus, the Jacobian Calabi--Yau 3-fold (\ref{Jacobian 3-fold in 4.1}) can have the singular fibers, only of types $III$, $I^*_0$ and $III^*$. This agrees with the results obtained in Section \ref{sssec4.2.1}. The monodromies of orders 2 and 4 characterize the non-Abelian gauge groups that form on F-theory compactifications of the Jacobian Calabi--Yau 3-folds (\ref{Jacobian 3-fold in 4.1}). 

\subsection{Mordell--Weil groups of some models, and models without $U(1)$ gauge field}
\label{ssec4.3}
We determine the Mordell--Weil groups of F-theory models on the Jacobian Calabi--Yau 3-folds (\ref{enhanced Jacobian 3-fold in 4.1}). We deduce that they do not have a $U(1)$ gauge field. 
\par We saw previously that the Weierstrass equation of the Jacobian Calabi--Yau 3-fold becomes (\ref{enhanced Jacobian 3-fold in 4.1}):
\begin{equation}
\label{enhanced Jacobian in 4.3}
\tau^2=\frac{1}{4}x^3-(t-\alpha_1)^3(t-\alpha_5)^3(t-\alpha_4)^2\cdot \Pi_{j=1}^8 (u-\beta_j)\, x,
\end{equation}
when the K3 fibers are most enhanced, namely when K3 fibers become the attractive K3 surface $S_{[2 \hspace{1mm} 0 \hspace{1mm} 2]}$ with the singularity type $E_7^2 D_4$. The Mordell--Weil group of this extremal K3 surface is known \cite{Nish, SZ}, and it is isomorphic to $\Z_2$. (See fibration no.4 in Table \ref{tablefibrations202 in A} in Appendix \ref{secA}.) Using an argument similar to those given in \cite{KCY4, Kdisc}, we consider the specialization to the K3 fiber to deduce that the Mordell--Weil group of the Jacobian Calabi--Yau 3-fold (\ref{enhanced Jacobian in 4.3}) is isomorphic to that of the K3 fiber. Therefore, we find that the Mordell--Weil group of the Jacobian Calabi--Yau 3-fold $J(Y)$ (\ref{enhanced Jacobian in 4.3}) is isomorphic to $\Z_2$:
\begin{equation}
MW (J(Y)) \cong \Z_2.
\end{equation}
Thus, the global structure \cite{AspinwallGross, AMrational, MMTW} of the gauge group forming on the 7-branes is given as follows:
\begin{equation}
E_7^2 \times SO(7) \times SU(2)^8 / \Z_2.
\end{equation}
The F-theory on the Jacobian Calabi--Yau 3-folds (\ref{enhanced Jacobian in 4.3}) does not have a $U(1)$ gauge field.

\subsection{Matter spectra}
\label{ssec4.4}
\par In this section, we deduce the candidate matter spectra on six-dimensional F-theory on the Jacobian Calabi--Yau 3-folds constructed in Section \ref{ssec4.1}. As we stated previously, 7-branes are wrapped on the discriminant components given by
\begin{eqnarray}
t & = & \alpha_i \hspace{5mm} (i=1, \ldots, 8) \\ \nonumber
u & = & \beta_j \hspace{5mm} (j=1, \ldots, 8).
\end{eqnarray}
7-branes wrapping the components are isomorphic to $\P^1$. 7-branes intersect at the points $(t,u)=(\alpha_i, \beta_j)$, $i,j=1, \ldots, 8$, in the base surface, and matter arises at these points. 
\par The base surface $B$ of the Jacobian Calabi--Yau 3-folds constructed in Section \ref{ssec4.1} is isomorphic to $\P^1\times \P^1$, $B\cong \P^1\times\P^1$, thus the number of tensor multiplets that arise in F-theory compactification on the Jacobian Calabi--Yau 3-folds is \cite{MV2}
\begin{eqnarray}
T & = & h^{1,1}(B=\P^1\times\P^1)-1 = 2-1 \\ \nonumber
& = & 1. 
\end{eqnarray}
The six-dimensional anomaly cancellation condition \cite{GSW6d, Sagnotti, Erler, Sch6d} then requires that
\begin{equation}
\label{anomaly cancellation in 4.4}
H-V=273-29=244.
\end{equation}
\par For simplicity, we only consider the case in which K3 fibers are most enhanced, namely K3 fibers become the extremal K3 surface with the singularity type $E_7^2 D_4$. This corresponds to the case where the coefficients $\alpha$ satisfy the relations (\ref{coeff in 4.1})
\begin{equation}
\label{coeff in 4.3}
\alpha_1=\alpha_2=\alpha_3, \hspace{5mm} \alpha_4=\alpha_8, \hspace{5mm} \alpha_5=\alpha_6=\alpha_7.
\end{equation}
As we saw in Section \ref{ssec4.3}, for this situation the Mordell--Weil rank of the Jacobian Calabi--Yau 3-folds is 0, and the Mordell--Weil group is isomorphic to $\Z_2$. 

\vspace{5mm}

\par First, we consider the case where a pair of $\beta$ coincide:
\begin{equation}
\beta_1=\beta_2.
\end{equation}
Then the equation of the Jacobian Calabi--Yau 3-fold (\ref{Jacobian 3-fold in 4.1}) becomes
\begin{equation}
\label{1st 3-fold in 4.4}
\tau^2=\frac{1}{4}x^3-(t-\alpha_1)^3(t-\alpha_5)^3(t-\alpha_4)^2\cdot (u-\beta_1)^2\cdot \Pi_{j=3}^8 (u-\beta_j)\cdot x
\end{equation}
and the discriminant is given as follows
\begin{equation}
\Delta \sim (t-\alpha_1)^9(t-\alpha_5)^9(t-\alpha_4)^6\cdot (u-\beta_1)^6\cdot \Pi_{j=3}^8 (u-\beta_j)^3.
\end{equation}
The non-Abelian gauge group forming in F-theory compactification is
\begin{equation}
E_7^2 \times SO(7)^2 \times SU(2)^6.
\end{equation}
Therefore, we have 
\begin{equation}
V=133\times 2+21\times 2+6\times 3=326.
\end{equation}
The anomaly cancellation condition (\ref{anomaly cancellation in 4.4}) requires that 
\begin{equation}
H=V+244=326+244=570.
\end{equation}
\par Matter arises at the 21 intersections 
\begin{equation}
(\alpha_i, \beta_j) \hspace{1cm} (i=1,4,5, \hspace{2mm} j=1,3,4,5,6,7,8).
\end{equation}
Parameters $\alpha_1, \alpha_4, \alpha_5$ can be sent to fixed values, e.g., 0, 1, $\infty$, under an automorphism of the first $\P^1$ in the base $\P^1\times\P^1$. Therefore, these are superfluous and are not actual parameters of the complex structure deformation. Among parameters of the complex structure deformation $\beta_1, \beta_3, \beta_4, \beta_5, \beta_6, \beta_7, \beta_8$, three of these can be fixed to specific values under an automorphism of the second $\P^1$ in the base $\P^1\times\P^1$. Thus, the number of the effective parameters of the complex structure deformation is four. The number of the neutral hypermultiplets arising from the complex structure deformations is therefore given by
\begin{equation}
H^0=1+4=5.
\end{equation}
Here, $H^0$ is used to denote the number of the neutral hypermultiplets. It follows that the sum of the dimensions of the representations of matter arising at the 21 intersections $(\alpha_i, \beta_j)$, $i=1,4, 5$, $j=1,3,4,5,6,7,8$, must be
\begin{equation}
570-5=565
\end{equation}
to cancel the anomaly.
\par $E_7$ angularity and $D_4$ singularity collide at two intersections $(t,u)=(\alpha_1, \beta_1), (\alpha_5, \beta_1)$, and $D_4$ singularities collide at the intersection $(t,u)=(\alpha_4, \beta_1)$. $E_7$ singularity and $A_1$ singularity collide at the 12 intersections $(t,u)=(\alpha_i, \beta_j)$, $i=1,5$, $j=3, \ldots, 8$. $D_4$ singularity and $A_1$ singularity collide at the six intersections $(t,u)=(\alpha_4, \beta_j)$, $j=3, \ldots, 8$. When the types of colliding singularities are fixed, from a symmetry argument, the representations of matter arising at the intersections at which the fixed types of singularities collide should be identical. 
\par If we assume that ${\bf 56}\oplus {\bf 7}\oplus {\bf 1}$ arises at the two intersections where $E_7$ angularity and $D_4$ singularity collide, ${\bf 21}\oplus {\bf 7}\oplus {\bf 1}$ arises at the intersection where $D_4$ singularities collide, $\frac{1}{2}{\bf 56}\oplus {\bf 2}$ arises at the 12 intersections where $E_7$ singularity and $A_1$ singularity collide and ${\bf 7}\oplus {\bf 1}$ arises at the six intersections where $D_4$ singularity and $A_1$ singularity collide, then the net dimension of the matter representations arising at the intersections of the 7-branes is
\begin{equation}
(56+7+1)\times 2+(21+7+1)+(28+2)\times 12+(7+1)\times 6=565.
\end{equation}
Therefore, these matter representations yield a candidate of consistent matter spectrum on F-theory compactification on the Calabi--Yau 3-fold (\ref{1st 3-fold in 4.4}) that satisfies the anomaly cancellation condition. Here $\frac{1}{2}{\bf 56}$ denotes the $\frac{1}{2}$-hypermultiplet of ${\bf 56}$ of $E_7$\footnote{The appearance of $\frac{1}{2}$-hypermultiplets of ${\bf 56}$ of $E_7$ in F-theory compactification was discussed in \cite{MV1, MV2, BIKMSV}. The base of elliptically fibered Jacobian Calabi--Yau 3-folds (\ref{1st 3-fold in 4.4}), $\P^1\times\P^1$, is isomorphic to Hirzebruch surface $\mathbb{F}_0$. Weierstrass coefficient $f$ of the Weierstrass equation $y^2=x^3+f\, x+g$ of Jacobian Calabi--Yau 3-fold (\ref{1st 3-fold in 4.4}) is given by $$f=(t-\alpha_1)^3(t-\alpha_5)^3(t-\alpha_4)^2\cdot (u-\beta_1)^2 \cdot \Pi^8_{j=3} (u-\beta_j).$$ Candidate $\frac{1}{2}$-hypermultiplets $\frac{1}{2}{\bf 56}$ localized at the twelve intersections have an interpretation as localized at the intersections of $E_7$ loci $t=\alpha_{1,5}$ and the zeroes of $\Pi^8_{j=3} (u-\beta_j)$ \cite{BIKMSV}.}. Having $\frac{1}{2}{\bf 56}\oplus {\bf 1}$ instead of $\frac{1}{2}{\bf 56}\oplus {\bf 2}$ at the 12 intersections $(t,u)=(\alpha_i, \beta_j)$, $i=1,5$, $j=3, \ldots, 8$, and ${\bf 7}\oplus {\bf 2}\oplus {\bf 1}$ instead of ${\bf 7}\oplus {\bf 1}$ at the six intersections $(t,u)=(\alpha_4, \beta_j)$, $j=3, \ldots, 8$ also yields another viable candidate of matter spectrum. Additionally, having matter representation ${\bf 56}\oplus {\bf 8}$, instead of ${\bf 56}\oplus {\bf 7}\oplus {\bf 1}$, at the intersections where $E_7$ angularity and $D_4$ singularity collide, or matter representation ${\bf 21}\oplus {\bf 8}$, instead of ${\bf 21}\oplus {\bf 7}\oplus {\bf 1}$, at the intersections where two $D_4$ singularities collide, also yield other viable candidate matter spectra. Furthermore, having matter representation ${\bf 56}\oplus {\bf 8}\oplus {\bf 1}$, instead of ${\bf 56}\oplus {\bf 7}\oplus {\bf 1}$, at the intersections where $E_7$ angularity and $D_4$ singularity collide, matter representation ${\bf 21}$, instead of ${\bf 21}\oplus {\bf 7}\oplus {\bf 1}$, at the intersection where two $D_4$ singularities collide, and matter representation ${\bf 7}\oplus {\bf 2}$, instead of ${\bf 7}\oplus {\bf 1}$, at the intersections where $D_4$ singularity and $A_1$ singularity collide also yields another viable candidate matter spectrum.
\par Because the Mordell--Weil group of the Calabi--Yau elliptic fibration (\ref{enhanced Jacobian 3-fold in 4.1}) is isomorphic to $\Z_2$, the global structure of the gauge group forming in F-theory compactification on the Calabi--Yau 3-fold (\ref{1st 3-fold in 4.4}) is
\begin{equation}
E_7^2 \times SO(7)^2 \times SU(2)^6 /\Z_2. 
\end{equation}

\vspace{5mm}

\par Next, we discuss the case where three pairs of the parameters $\beta$ coincide:
\begin{eqnarray}
\beta_1=\beta_2 \\ \nonumber
\beta_3=\beta_4 \\ \nonumber
\beta_5=\beta_6.
\end{eqnarray}
The equation of the Calabi--Yau 3-fold (\ref{enhanced Jacobian 3-fold in 4.1}) becomes
\begin{equation}
\label{2nd 3-fold in 4.4}
\tau^2=\frac{1}{4}x^3-(t-\alpha_1)^3(t-\alpha_5)^3(t-\alpha_4)^2\cdot (u-\beta_1)^2(u-\beta_3)^2(u-\beta_5)^2\cdot \Pi_{j=7}^8 (u-\beta_j)\cdot x
\end{equation}
and the discriminant is given as follows
\begin{equation}
\Delta \sim (t-\alpha_1)^9(t-\alpha_5)^9(t-\alpha_4)^6\cdot (u-\beta_1)^6(u-\beta_3)^6(u-\beta_5)^6\cdot \Pi_{j=7}^8 (u-\beta_j)^3.
\end{equation}
Non-Abelian gauge group forming in F-theory compactification is:
\begin{equation}
E_7^2 \times SO(7)^4 \times SU(2)^2.
\end{equation}
Thus, 
\begin{equation}
V=133\times 2+21\times 4+3\times 2=356.
\end{equation}
Anomaly cancellation condition (\ref{anomaly cancellation in 4.4}) requires that 
\begin{equation}
H=356+244=600.
\end{equation}
Among the parameters of the complex structure deformation, $\beta_1, \beta_3, \beta_5, \beta_7, \beta_8$, three can be sent to fixed values under an automorphism of $\P^1$. Therefore, the actual number of the parameters of the complex structure deformation is two. From this, we obtain the number of the neutral hypermultiplets arising from the complex structure deformations as
\begin{equation}
H^0=1+2=3.
\end{equation}
7-branes intersect in fifteen points at $(t,u)=(\alpha_i, \beta_j)$, $i=1,4,5$, $j=1,3,5,7,8$, and matter arises at these intersections. The net representation dimensions of matter arising from these intersections must be 
\begin{equation}
600-3=597
\end{equation}
owing to the anomaly cancellation condition. 
\par $E_7$ angularity and $D_4$ singularity collide at six intersections $(t,u)=(\alpha_i, \beta_j)$, $i=1,5$, $j=1,3,5$, and $D_4$ singularities collide at the three intersections $(t,u)=(\alpha_4, \beta_j)$, $j=1,3,5$. $E_7$ singularity and $A_1$ singularity collide at the four intersections $(t,u)=(\alpha_i, \beta_j)$, $i=1,5$, $j=7,8$. $D_4$ singularity and $A_1$ singularity collide at the two intersections $(t,u)=(\alpha_4, \beta_j)$, $j=7,8$.
\par We assume that ${\bf 56}\oplus {\bf 7}\oplus {\bf 1}$ arises at the six intersections where $E_7$ angularity and $D_4$ singularity collide, ${\bf 21}\oplus {\bf 7}\oplus {\bf 1}$ arises at the three intersection where $D_4$ singularities collide, $\frac{1}{2}{\bf 56}$ arises at the four intersections where $E_7$ singularity and $A_1$ singularity collide and ${\bf 7}$ arises at the two intersections where $D_4$ singularity and $A_1$ singularity collide, then the net dimension of the matter representations arising at the intersections of the 7-branes is
\begin{equation}
(56+7+1)\times 6+(21+7+1)\times 3+28\times 4+7\times 2=597.
\end{equation}
Thus, this yields a consistent matter candidate on F-theory on the Jacobian Calabi--Yau 3-fold (\ref{2nd 3-fold in 4.4}). Having ${\bf 56}\oplus {\bf 7}$, instead of ${\bf 56}\oplus {\bf 7}\oplus {\bf 1}$, at the six intersections where $E_7$ angularity and $D_4$ singularity collide, and ${\bf 7}\oplus {\bf 2}\oplus {\bf 1}$, instead of ${\bf 7}$, at the two intersections where $D_4$ singularity and $A_1$ singularity collide, also yields another viable candidate matter spectrum. Furthermore, having ${\bf 56}\oplus {\bf 7}$, instead of ${\bf 56}\oplus {\bf 7}\oplus {\bf 1}$, at the six intersections where $E_7$ angularity and $D_4$ singularity collide, $\frac{1}{2}{\bf 56}\oplus {\bf 1}$, instead of $\frac{1}{2}{\bf 56}$, at the four intersections where $E_7$ singularity and $A_1$ singularity collide and ${\bf 7}\oplus {\bf 1}$, instead of ${\bf 7}$, at the two intersections where $D_4$ singularity and $A_1$ singularity collide, yields another viable candidate matter spectrum. In addition to these, having ${\bf 56}\oplus {\bf 8}\oplus {\bf 1}\oplus {\bf 1}$, instead of ${\bf 56}\oplus {\bf 7}\oplus {\bf 1}$, at the six intersections where $E_7$ angularity and $D_4$ singularity collide, ${\bf 21}$, instead of ${\bf 21}\oplus {\bf 7} \oplus {\bf 1}$, at the three intersections where two $D_4$ singularities collide, $\frac{1}{2}{\bf 56}\oplus {\bf 2}$, instead of $\frac{1}{2}{\bf 56}$, at the four intersections where $E_7$ singularity and $A_1$ singularity collide and ${\bf 7}\oplus {\bf 2}$, instead of ${\bf 7}$, at the two intersections where $D_4$ singularity and $A_1$ singularity collide, yields another viable candidate matter spectrum.
\par The global structure of the gauge group forming in F-theory compactification on the Calabi--Yau 3-fold (\ref{2nd 3-fold in 4.4}) is
\begin{equation}
E_7^2 \times SO(7)^4 \times SU(2)^2 /\Z_2. 
\end{equation}

\vspace{5mm}

\par We then consider the case a triplet of parameters $\beta$ coincide:
\begin{equation}
\beta_1=\beta_2=\beta_3.
\end{equation}
In this situation, the equation of the Calabi--Yau 3-fold (\ref{enhanced Jacobian 3-fold in 4.1}) becomes
\begin{equation}
\label{3rd 3-fold in 4.4}
\tau^2=\frac{1}{4}x^3-(t-\alpha_1)^3(t-\alpha_5)^3(t-\alpha_4)^2\cdot (u-\beta_1)^3\cdot \Pi_{j=4}^8 (u-\beta_j)\cdot x
\end{equation}
and the discriminant is given as follows
\begin{equation}
\Delta \sim (t-\alpha_1)^9(t-\alpha_5)^9(t-\alpha_4)^6\cdot (u-\beta_1)^9\cdot \Pi_{j=4}^8 (u-\beta_j)^3.
\end{equation}
Non-Abelian gauge group forming in F-theory compactification is:
\begin{equation}
E_7^3 \times SO(7) \times SU(2)^5.
\end{equation}
Therefore, we have
\begin{equation}
133\times 3+21+3\times 5=435.
\end{equation}
The anomaly cancellation condition requires that
\begin{equation}
H=435+244=679.
\end{equation}
Among the parameters of the complex structure deformation $\beta_1, \beta_4, \beta_5, \beta_6, \beta_7, \beta_8$, three can be sent to fixed values under an automorphism of $\P^1$. Therefore, the number of the effective parameters of the complex structure deformation is three. From this, we obtain the number of the neutral hypermultiplets arising from the complex structure deformations as
\begin{equation}
H^0=1+3=4.
\end{equation}
7-branes intersect in eighteen points at $(t,u)=(\alpha_i, \beta_j)$, $i=1,4,5$, $j=1,4,5,6,7,8$. The net representation dimensions of matter arising from these intersections must be 
\begin{equation}
679-4=675
\end{equation}
owing to the anomaly cancellation condition.
\par $E_7$ angularities collide at the two intersections $(t,u)=(\alpha_i, \beta_1)$, $i=1,5$, and $D_4$ singularity and $E_7$ singularity collide at the intersection $(t,u)=(\alpha_4, \beta_1)$. $E_7$ singularity and $A_1$ singularity collide at the ten intersections $(t,u)=(\alpha_i, \beta_j)$, $i=1,5$, $j=4, \cdots, 8$. $D_4$ singularity and $A_1$ singularity collide at the five intersections $(t,u)=(\alpha_4, \beta_j)$, $j=4, \cdots ,8$.
\par We assume that ${\bf 133}$ arises at the two intersections where two $E_7$ angularities collide, ${\bf 56}\oplus {\bf 7}\oplus {\bf 1}$ arises at the intersection where $D_4$ and $E_7$ singularities collide, $\frac{1}{2}{\bf 56}\oplus {\bf 2}$ arises at the ten intersections where $E_7$ singularity and $A_1$ singularity collide and ${\bf 7}\oplus {\bf 2}$ arises at the five intersections where $D_4$ singularity and $A_1$ singularity collide, then the net dimension of the matter representations arising at the intersections of the 7-branes is
\begin{equation}
133\times 2+(56+7+1)+(28+2)\times 10+(7+2)\times 5=675.
\end{equation}
Thus, the anomaly cancels and this yields a consistent matter candidate on F-theory on the Jacobian Calabi--Yau 3-fold (\ref{3rd 3-fold in 4.4}). Having matter representation ${\bf 56}\oplus {\bf 8}$, instead of ${\bf 56}\oplus {\bf 7}\oplus {\bf 1}$, at the intersection where $E_7$ angularity and $D_4$ singularity collide also yields another viable candidate matter spectrum.
\par The global structure of the gauge group forming in F-theory compactification on the Calabi--Yau 3-fold (\ref{3rd 3-fold in 4.4}) is
\begin{equation}
E_7^3 \times SO(7) \times SU(2)^5 /\Z_2. 
\end{equation}

\vspace{5mm}

\par We now consider the case a triplet and a pair of parameters $\beta$ coincide: 
\begin{eqnarray}
\beta_1=\beta_2=\beta_3 \\ \nonumber
\beta_4=\beta_8.
\end{eqnarray}
In this situation, the equation of the Jacobian Calabi--Yau 3-fold (\ref{enhanced Jacobian 3-fold in 4.1}) becomes:
\begin{equation}
\label{4th 3-fold in 4.4}
\tau^2=\frac{1}{4}x^3-(t-\alpha_1)^3(t-\alpha_5)^3(t-\alpha_4)^2\cdot (u-\beta_1)^3(u-\beta_4)^2\cdot \Pi_{j=5}^7 (u-\beta_j)\cdot x.
\end{equation}
The discriminant is given as follows
\begin{equation}
\Delta \sim (t-\alpha_1)^9(t-\alpha_5)^9(t-\alpha_4)^6\cdot (u-\beta_1)^9(u-\beta_4)^6\cdot \Pi_{j=5}^7 (u-\beta_j)^3.
\end{equation}
Non-Abelian gauge group forming in F-theory compactification is:
\begin{equation}
E_7^3 \times SO(7)^2 \times SU(2)^3.
\end{equation}
Therefore, we have
\begin{equation}
133\times 3+21\times 2+3\times 3=450.
\end{equation}
The anomaly cancellation condition requires that
\begin{equation}
H=450+244=694.
\end{equation}
Among the parameters of the complex structure deformation $\beta_1, \beta_4, \beta_5, \beta_6, \beta_7$, three can be sent to fixed values under an automorphism of $\P^1$. Therefore, the number of the effective parameters of the complex structure deformation is two. From this, we obtain the number of the neutral hypermultiplets arising from the complex structure deformations as
\begin{equation}
H^0=1+2=3.
\end{equation}
7-branes intersect in fifteen points at $(t,u)=(\alpha_i, \beta_j)$, $i=1,4,5$, $j=1,4,5,6,7$. The net representation dimensions of matter arising from these intersections must be 
\begin{equation}
694-3=691
\end{equation}
owing to the anomaly cancellation condition.
\par $E_7$ angularities collide at the two intersections $(t,u)=(\alpha_i, \beta_1)$, $i=1,5$, and $D_4$ singularity and $E_7$ singularity collide at the three intersection points $(t,u)=(\alpha_4, \beta_1), (\alpha_i, \beta_4)$, $i=1,5$. $D_4$ singularities collide at the intersection point $(t,u)=(\alpha_4, \beta_4)$. $E_7$ singularity and $A_1$ singularity collide at the six intersections $(t,u)=(\alpha_i, \beta_j)$, $i=1,5$, $j=5,6,7$. $D_4$ singularity and $A_1$ singularity collide at the three intersections $(t,u)=(\alpha_4, \beta_j)$, $j=5,6,7$.
\par We assume that ${\bf 133}$ arises at the two intersections where two $E_7$ angularities collide, ${\bf 56}\oplus {\bf 7}\oplus {\bf 1}$ arises at the three intersection points where $D_4$ and $E_7$ singularities collide, ${\bf 21}\oplus {\bf 7}\oplus {\bf 1}$ arises at the intersection point where two $D_4$ singularities collide, $\frac{1}{2}{\bf 56}\oplus {\bf 2}$ arises at the six intersections where $E_7$ singularity and $A_1$ singularity collide and ${\bf 7}\oplus {\bf 1}$ arises at the three intersections where $D_4$ singularity and $A_1$ singularity collide, then the net dimension of the matter representations arising at the intersections of the 7-branes is
\begin{equation}
133\times 2+(56+7+1)\times 3+(21+7+1)+(28+2)\times 6+(7+1)\times 3=691.
\end{equation}
Thus, the anomaly cancels and this yields a consistent matter candidate on F-theory on the Jacobian Calabi--Yau 3-fold (\ref{4th 3-fold in 4.4}). Having matter representation ${\bf 56}\oplus {\bf 7}$, instead of ${\bf 56}\oplus {\bf 7}\oplus {\bf 1}$, at the intersections where $E_7$ angularity and $D_4$ singularity collide, and matter representation ${\bf 7}\oplus {\bf 2}$, instead of ${\bf 7}\oplus {\bf 1}$, at the intersections where $D_4$ and $A_1$ singularities collide also yields another viable candidate matter spectrum. Furthermore, having matter representation ${\bf 133}\oplus {\bf 1}$, instead of ${\bf 133}$, at the intersections where two $E_7$ angularities collide, matter representation ${\bf 56}\oplus {\bf 8}\oplus {\bf 1}$, instead of ${\bf 56}\oplus {\bf 7}\oplus {\bf 1}$, at the intersections where $E_7$ angularity and $D_4$ singularity collide, matter representation ${\bf 21}$, instead of ${\bf 21}\oplus {\bf 7}\oplus {\bf 1}$, at the intersection where two $D_4$ singularities collide, and matter representation ${\bf 7}\oplus {\bf 2}$, instead of ${\bf 7}\oplus {\bf 1}$, at the intersections where $D_4$ and $A_1$ singularities collide yields another viable candidate matter spectrum.
\par The global structure of the gauge group forming in F-theory compactification on the Calabi--Yau 3-fold (\ref{4th 3-fold in 4.4}) is
\begin{equation}
E_7^3 \times SO(7)^2 \times SU(2)^3 /\Z_2. 
\end{equation}

\vspace{5mm}

\par We consider the case a triplet and two pairs of parameters $\beta$ coincide:
\begin{eqnarray}
\beta_1=\beta_2=\beta_3 \\ \nonumber
\beta_4=\beta_8 \\ \nonumber
\beta_5=\beta_6.
\end{eqnarray}
In this situation, the equation of the Jacobian Calabi--Yau 3-fold (\ref{enhanced Jacobian 3-fold in 4.1}) becomes:
\begin{equation}
\label{5th 3-fold in 4.4}
\tau^2=\frac{1}{4}x^3-(t-\alpha_1)^3(t-\alpha_5)^3(t-\alpha_4)^2\cdot (u-\beta_1)^3(u-\beta_4)^2(u-\beta_5)^2(u-\beta_7)\cdot x.
\end{equation}
The discriminant is given as follows
\begin{equation}
\Delta \sim (t-\alpha_1)^9(t-\alpha_5)^9(t-\alpha_4)^6\cdot (u-\beta_1)^9(u-\beta_4)^6(u-\beta_5)^6(u-\beta_7)^3.
\end{equation}
Non-Abelian gauge group forming in F-theory compactification is:
\begin{equation}
E_7^3 \times SO(7)^3 \times SU(2).
\end{equation}
Therefore, we have
\begin{equation}
133\times 3+21\times 3+3=465.
\end{equation}
The anomaly cancellation condition requires that
\begin{equation}
H=465+244=709.
\end{equation}
Among the parameters of the complex structure deformation $\beta_1, \beta_4, \beta_5, \beta_7$, three can be sent to fixed values under an automorphism of $\P^1$. Therefore, the number of the effective parameters of the complex structure deformation is one. From this, we obtain the number of the neutral hypermultiplets as
\begin{equation}
H^0=1+1=2.
\end{equation}
7-branes intersect in twelve points at $(t,u)=(\alpha_i, \beta_j)$, $i=1,4,5$, $j=1,4,5,7$. The net representation dimensions of matter arising from these intersections must be 
\begin{equation}
709-2=707
\end{equation}
owing to the anomaly cancellation condition.
\par $E_7$ angularities collide at the two intersections $(t,u)=(\alpha_i, \beta_1)$, $i=1,5$, and $E_7$ singularity and $D_4$ singularity collide at the five intersection points $(t,u)=(\alpha_i, \beta_j)$, $i=1,5, j=4,5$, $(\alpha_4, \beta_1)$. $D_4$ singularities collide at the two intersection points $(t,u)=(\alpha_4, \beta_j)$, $j=4,5$. $E_7$ singularity and $A_1$ singularity collide at the two intersections $(t,u)=(\alpha_i, \beta_7)$, $i=1,5$. $D_4$ singularity and $A_1$ singularity collide at the intersection $(t,u)=(\alpha_4, \beta_7)$.
\par We assume that ${\bf 133}$ arises at the two intersections where two $E_7$ angularities collide, ${\bf 56}\oplus {\bf 7}$ arises at the five intersection points where $E_7$ and $D_4$ singularities collide, ${\bf 21}\oplus {\bf 7}\oplus {\bf 1}$ arises at the two intersection points where two $D_4$ singularities collide, $\frac{1}{2}{\bf 56}\oplus {\bf 2}$ arises at the two intersections where $E_7$ singularity and $A_1$ singularity collide, and ${\bf 7}\oplus {\bf 1}$ arises at the intersection where $D_4$ singularity and $A_1$ singularity collide, then the net dimension of the matter representations arising at the intersections of the 7-branes is
\begin{equation}
133\times 2+(56+7)\times 5+(21+7+1)\times 2+(28+2)\times 2+(7+1)=707.
\end{equation}
Thus, the anomaly cancels and this yields a consistent matter candidate on F-theory on the Jacobian Calabi--Yau 3-fold (\ref{5th 3-fold in 4.4}). Having matter representation ${\bf 56}\oplus {\bf 7}\oplus {\bf 1}$, instead of ${\bf 56}\oplus {\bf 7}$, at the intersections where $E_7$ angularity and $D_4$ singularity collide, matter representation $\frac{1}{2}{\bf 56}$, instead of $\frac{1}{2}{\bf 56}\oplus {\bf 2}$, at the intersections where $E_7$ and $A_1$ singularities collide, and matter representation ${\bf 7}$, instead of ${\bf 7}\oplus {\bf 1}$, at the intersection where $D_4$ and $A_1$ singularities collide also yields another viable candidate matter spectrum. Furthermore, having matter representation ${\bf 133}\oplus {\bf 1}$, instead of ${\bf 133}$, at the intersections where two $E_7$ angularities collide, and matter representation $\frac{1}{2}{\bf 56}\oplus {\bf 1}$, instead of $\frac{1}{2}{\bf 56}\oplus {\bf 2}$, at the intersections where $E_7$ and $A_1$ singularities collide yields another viable candidate matter spectrum. Additionally, having matter representation ${\bf 56}\oplus {\bf 8}\oplus {\bf 1}\oplus {\bf 1}$, instead of ${\bf 56}\oplus {\bf 7}$, at the intersections where $E_7$ angularity and $D_4$ singularity collide, matter representation ${\bf 21}$, instead of ${\bf 21}\oplus {\bf 7}\oplus {\bf 1}$, at the intersections where two $D_4$ singularities collide, and matter representation ${\bf 7}\oplus {\bf 2}$, instead of ${\bf 7}\oplus {\bf 1}$, at the intersection where $D_4$ and $A_1$ singularities collide also yields another viable candidate matter spectrum.
\par The global structure of the gauge group forming in F-theory compactification on the Calabi--Yau 3-fold (\ref{5th 3-fold in 4.4}) is
\begin{equation}
E_7^3 \times SO(7)^3 \times SU(2) /\Z_2. 
\end{equation}

\vspace{5mm}

\par We finally discuss the case two triplets and a pair of parameters $\beta$ coincide:
\begin{eqnarray}
\beta_1=\beta_2=\beta_3 \\ \nonumber
\beta_4=\beta_8 \\ \nonumber
\beta_5=\beta_6=\beta_7.
\end{eqnarray}
In this situation, the equation of the Jacobian Calabi--Yau 3-fold (\ref{enhanced Jacobian 3-fold in 4.1}) becomes:
\begin{equation}
\label{6th 3-fold in 4.4}
\tau^2=\frac{1}{4}x^3-(t-\alpha_1)^3(t-\alpha_5)^3(t-\alpha_4)^2\cdot (u-\beta_1)^3(u-\beta_5)^3(u-\beta_4)^2\cdot x.
\end{equation}
The discriminant is given as follows
\begin{equation}
\Delta \sim (t-\alpha_1)^9(t-\alpha_5)^9(t-\alpha_4)^6\cdot (u-\beta_1)^9(u-\beta_5)^9(u-\beta_4)^6.
\end{equation}
Non-Abelian gauge group forming in F-theory compactification is:
\begin{equation}
E_7^4 \times SO(7)^2.
\end{equation}
Thus, we have
\begin{equation}
133\times 4+21\times 2=574.
\end{equation}
The anomaly cancellation condition requires that
\begin{equation}
H=574+244=818.
\end{equation}
The parameters of the complex structure deformation, $\beta_1, \beta_4, \beta_5$, can be sent to fixed values under an automorphism of $\P^1$. Therefore, the number of the effective parameters of the complex structure deformation is zero, and the complex structure is fixed for this situation. From this, we obtain the number of the neutral hypermultiplets as
\begin{equation}
H^0=1+0=1.
\end{equation}
7-branes intersect in nine points at $(t,u)=(\alpha_i, \beta_j)$, $i=1,4,5$, $j=1,4,5$. The net representation dimensions of matter arising from these intersections must be 
\begin{equation}
818-1=817
\end{equation}
owing to the anomaly cancellation condition.
\par $E_7$ angularities collide at the four intersections $(t,u)=(\alpha_i, \beta_j)$, $i=1,5$, $j=1,5$, and $E_7$ singularity and $D_4$ singularity collide at the four intersection points $(t,u)=(\alpha_4, \beta_j)$, $j=1,5$, $(\alpha_i, \beta_4)$, $i=1,5$. $D_4$ singularities collide at the intersection point $(t,u)=(\alpha_4, \beta_4)$.
\par We assume that ${\bf 133}$ arises at the four intersections where two $E_7$ angularities collide, ${\bf 56}\oplus {\bf 7}\oplus {\bf 1}$ arises at the four intersection points where $E_7$ and $D_4$ singularities collide, and ${\bf 21}\oplus {\bf 7}\oplus {\bf 1}$ arises at the intersection point where two $D_4$ singularities collide, then the net dimension of the matter representations arising at the intersections of the 7-branes is
\begin{equation}
133\times 4+(56+7+1)\times 4+(21+7+1)=817.
\end{equation}
Thus, the anomaly cancels and this yields a consistent matter candidate on F-theory on the Jacobian Calabi--Yau 3-fold (\ref{6th 3-fold in 4.4}). Having matter representation ${\bf 56}\oplus {\bf 8}$, instead of ${\bf 56}\oplus {\bf 7}\oplus {\bf 1}$, at the intersections where $E_7$ angularity and $D_4$ singularity collide, or matter representation ${\bf 21}\oplus {\bf 8}$, instead of ${\bf 21}\oplus {\bf 7}\oplus {\bf 1}$, at the intersection where two $D_4$ singularities collide, also yield other viable candidate matter spectra. Additionally, having matter representation ${\bf 133}\oplus {\bf 1}$, instead of ${\bf 133}$, at the intersections where two $E_7$ angularities collide, and matter representation ${\bf 56}\oplus {\bf 7}$, instead of ${\bf 56}\oplus {\bf 7}\oplus {\bf 1}$, at the intersections where $E_7$ and $D_4$ singularities collide, also yields another viable candidate matter spectrum. Furthermore, having matter representation ${\bf 133}\oplus {\bf 1}$, instead of ${\bf 133}$, at the intersections where two $E_7$ angularities collide, matter representation ${\bf 56}\oplus {\bf 8}\oplus {\bf 1}$, instead of ${\bf 56}\oplus {\bf 7}\oplus {\bf 1}$, at the intersections where $E_7$ and $D_4$ singularities collide, and matter representation ${\bf 21}$, instead of ${\bf 21}\oplus {\bf 7}\oplus {\bf 1}$, at the intersection where two $D_4$ singularities collide, yields another viable candidate matter spectrum.
\par The global structure of the gauge group forming in F-theory compactification on the Calabi--Yau 3-fold (\ref{6th 3-fold in 4.4}) is
\begin{equation}
E_7^4 \times SO(7)^2 /\Z_2. 
\end{equation}

\vspace{1cm}

\par We have deduced candidate matter spectra on F-theory on the constructed elliptically fibered Calabi--Yau 3-folds. We have observed that it is natural to expect that ${\bf 133}$ (or ${\bf 133}\oplus {\bf 1}$) arises at the collision of two $E_7$ singularities \footnote{The authors of \cite{AHMT1803} discussed the collision of two $E_7$ singularities in the context of four-dimensional conformal matter in F-theory.}, to cancel the anomaly\footnote{A similar observation was made for the collision of two $E_6$ singularities for elliptically fibered Calabi--Yau 3-folds of ``Fermat-types'' in \cite{Kimura1810}, where it was argued that ${\bf 78}\oplus {\bf 1}$ is expected to arise at this intersection.}. Under this assumption, we observed that matter representation arising at the collision of $E_7$ and $A_1$ singularities should include the $\frac{1}{2}$-hypermultiplet $\frac{1}{2}{\bf 56}$ of $E_7$. If matter representation at this intersection includes ${\bf 56}$, the net dimension of matter representations exceeds the number required by the anomaly cancellation condition by a large amount. There appear a few possibilities for matter representation at the collision of $E_7$ and $D_4$ singularities, such as matter representations ${\bf 56}\oplus {\bf 7}\oplus {\bf 1}$ or  ${\bf 56}\oplus {\bf 8}$, and whether to include ${\bf 1}$. We observed a few possibilities for matter representation at the collision of two $D_4$ singularities, such as ${\bf 21}\oplus {\bf 7}\oplus {\bf 1}$, ${\bf 21}\oplus {\bf 8}$, or ${\bf 21}$. There are also a few possibilities for matter at the collision of $E_7$ and $A_1$ singularities, such as $\frac{1}{2}{\bf 56}$, $\frac{1}{2}{\bf 56}\oplus {\bf 1}$, or $\frac{1}{2}{\bf 56} \oplus {\bf 2}$, and similarly for the collision of $D_4$ and $A_1$ singularities.

\section{Conclusions}
\label{sec5}
In this study, we have investigated the points in the eight-dimensional moduli of non-geometric heterotic strings with unbroken $\mathfrak{e}_7\mathfrak{e}_7$ algebra, at which the ranks of the non-Abelian gauge groups on the F-theory side are enhanced to 18. The gauge groups at these points do not allow for the perturbative interpretation on the heterotic side. We demonstrated in this study that these theories can be seen as deformations of the stable degenerations owing to an effect of coincident 7-branes. This effect corresponds to the insertion of 5-branes from the heterotic viewpoint. We also discussed the application to $SO(32)$ heterotic strings. 
\par K3 surfaces on the F-theory side of the moduli become extremal, when the non-Abelian gauge groups are enhanced to rank 18. We studied the Weierstrass equations of the extremal K3 surfaces that appear on the F-theory side of the eight-dimensional moduli of non-geometric heterotic strings with unbroken $\mathfrak{e}_7\mathfrak{e}_7$. The points in the moduli at which the ranks of the non-Abelian gauge groups are enhanced to 17 on the F-theory side also do not allow for the perturbative interpretations of the gauge groups on the heterotic side. It can be interesting to study these points in the moduli, and this can be a direction of future study.
\par We have also built elliptically fibered Calabi--Yau 3-folds, by fibering an elliptic K3 surface, which belongs to the F-theory side of the eight-dimensional moduli of non-geometric heterotic strings with unbroken $\mathfrak{e}_7\mathfrak{e}_7$ algebra, over $\P^1$. We analyzed six-dimensional F-theory compactifications on the built elliptic Calabi--Yau 3-folds. When we tune the parameters for the defining equations of these elliptic Calabi--Yau 3-folds, highly enhanced gauge groups form on the 7-branes. Eight-dimensional F-theory compactified on the extremal K3 fibers $S_{[2 \hspace{1mm} 0 \hspace{1mm} 2]}$ of these specific tuned Calabi--Yau spaces has non-geometric heterotic duals. Determining whether this duality extends to six-dimensional theories, namely whether F-theory compactifications on the total Calabi--Yau 3-folds have dual non-geometric heterotic strings, is a likely direction of future study.
\par We have also deduced viable candidate matter spectra on F-theory on the constructed elliptically fibered Calabi--Yau 3-folds, for the case when K3 fibers are most enhanced (\ref{coeff in 4.1}). There are certain ambiguities such as whether matter arising at intersections of 7-branes includes ${\bf 1}$, or ${\bf 2}$, or does not include these. There are also ambiguities of whether ${\bf 7}\oplus {\bf 1}$ or ${\bf 8}$ arises where $D_4$ singularities and $D_4$ and $A_1$ singularities collide. There is also an ambiguity of whether matter arising at the collision of $D_4$ singularities includes ${\bf 7}\oplus {\bf 1}$, or ${\bf 8}$, or does not include these. Except for these ambiguities, the possibility of candidate matter spectra appears unique. We have observed that either ${\bf 133}$ or ${\bf 133}\oplus {\bf 1}$ arises where two $E_7$ singularities collide. We have also observed that matter arising where $E_7$ and $A_1$ singularities collide should include the $\frac{1}{2}$-hypermultiplet, $\frac{1}{2}{\bf 56}$ of $E_7$, to cancel the anomaly. Confirming the actual matter spectra by analyzing the resolution of the singularities of the Jacobian Calabi--Yau 3-folds is also a likely direction of future study.

\section*{Acknowledgments}

We would like to thank Shun'ya Mizoguchi for discussions. This work is partially supported by Grant-in-Aid for Scientific Research {\#}16K05337 from the Ministry of Education, Culture, Sports, Science and Technology of Japan.

\newpage
\appendix
\section{Elliptic fibrations of attractive K3 $S_{[2 \hspace{1mm} 0 \hspace{1mm} 2]}$}
\label{secA}
\cite{Nish} classified the types of the elliptic fibrations of the attractive K3 surface with the discriminant four, $S_{[2 \hspace{1mm} 0 \hspace{1mm} 2]}$, and computed the Mordell--Weil groups of the fibrations. We present in Table \ref{tablefibrations202 in A} the types of the elliptic fibrations and the Mordell--Weil groups of the attractive K3 surface $S_{[2 \hspace{1mm} 0 \hspace{1mm} 2]}$ determined in \cite{Nish}.

\begingroup
\renewcommand{\arraystretch}{1.1}
\begin{table}[htb]
\centering
  \begin{tabular}{|c|c|c|} \hline
$
\begin{array}{c}
\mbox{Elliptic fibrations of}\\
\mbox{$S_{[2 \hspace{1mm} 0 \hspace{1mm} 2]}$} 
\end{array}
$ & type of singularity & MW group \\ \hline
No.1 & $E^2_8 A^2_1$ & 0 \\ 
No.2 &  $E_8 D_{10}$ & 0 \\
No.3 & $D_{16} A^2_1$ & $\Z_2$ \\
No.4 & $E^2_7 D_4$ & $\Z_2$ \\ 
No.5 & $E_7 D_{10} A_1$ & $\Z_2$ \\
No.6 & $A_{17} A_1$ & $\Z_3$ \\ 
No.7 & $D_{18}$ & 0 \\ 
No.8 &  $D_{12} D_6$ & $\Z_2$ \\
No.9 & $D^2_8 A^2_1$ & $\Z_2 \oplus \Z_2$ \\
No.10 & $A_{15} A_3$ & $\Z_4$ \\ 
No.11 & $E_6 A_{11}$ & $\Z\oplus \Z_3$ \\
No.12 & $D^3_6$ & $\Z_2\oplus\Z_2$ \\ 
No.13 & $A^2_9$ & $\Z_5$ \\ \hline
\end{tabular}
\caption{\label{tablefibrations202 in A}List of the singularity types of the elliptic fibrations of K3 surface $S_{[2 \hspace{1mm} 0 \hspace{1mm} 2]}$, and the Mordell--Weil groups of the fibrations.}
\end{table}
\endgroup

\section{Types of the singular fibers of extremal rational elliptic surfaces}
\label{secB}
The types of the singular fibers of the extremal rational elliptic surfaces \cite{MP} are presented in Table \ref{tablelisttypes extremalRES in B}. The complex structures of the extremal rational elliptic surfaces are uniquely specified by the types of the singular fibers, except rational elliptic surfaces with two type $I^*_0$ fibers, $X_{[0^*, \hspace{1mm} 0^*]}(j)$ \cite{MP}.

\begingroup
\renewcommand{\arraystretch}{1.5}
\begin{table}[htb]
\centering
  \begin{tabular}{|c|c|c|} \hline
$
\begin{array}{c}
\mbox{Extremal rational}\\
\mbox{elliptic surface} 
\end{array}
$ & $
\begin{array}{c}
\mbox{Type of}\\
\mbox{singular fiber} 
\end{array}
$ & $
\begin{array}{c}
\mbox{Type of}\\
\mbox{singularity} 
\end{array}
$ \\ \hline
$X_{[II, \hspace{1mm} II^*]}$ & $II$, $II^*$ & $E_8$  \\ \hline
$X_{[III, \hspace{1mm} III^*]}$ & $III$, $III^*$ & $E_7A_1$  \\ \hline
$X_{[IV, \hspace{1mm} IV^*]}$ & $IV$, $IV^*$ & $E_6A_2$ \\ \hline
$X_{[0^*, \hspace{1mm} 0^*]}(j)$ & $I^*_0$, $I^*_0$ & $D_4^2$ \\ \hline
$X_{[II^*, \hspace{1mm} 1, 1]}$ & $II^*$ $I_1$ $I_1$ & $E_8$ \\ \hline
$X_{[III^*, \hspace{1mm} 2, 1]}$ & $III^*$ $I_2$ $I_1$ & $E_7A_1$ \\ \hline
$X_{[IV^*, \hspace{1mm} 3, 1]}$ & $IV^*$ $I_3$ $I_1$ & $E_6A_2$ \\ \hline
$X_{[4^*, \hspace{1mm} 1, 1]}$ & $I_4^*$ $I_1$ $I_1$ & $D_8$ \\ \hline
$X_{[2^*, \hspace{1mm} 2, 2]}$ & $I^*_2$ $I_2$ $I_2$ & $D_6A_1^2$ \\ \hline
$X_{[1^*, \hspace{1mm} 4, 1]}$ & $I_1^*$ $I_4$ $I_1$ & $D_5A_3$ \\ \hline
$X_{[9, 1, 1, 1]}$ & $I_9$ $I_1$ $I_1$ $I_1$ & $A_8$ \\ \hline
$X_{[8, 2, 1, 1]}$ & $I_8$ $I_2$ $I_1$ $I_1$ & $A_7A_1$ \\ \hline 
$X_{[6, 3, 2, 1]}$ & $I_6$ $I_3$ $I_2$ $I_1$ & $A_5A_2A_1$ \\ \hline
$X_{[5, 5, 1, 1]}$ & $I_5$ $I_5$ $I_1$ $I_1$ & $A_4^2$  \\ \hline
$X_{[4, 4, 2, 2]}$ & $I_4$ $I_4$ $I_2$ $I_2$ & $A_3^2A_1^2$ \\ \hline
$X_{[3, 3, 3, 3]}$ & $I_3$ $I_3$ $I_3$ $I_3$ & $A_2^4$ \\ \hline
\end{tabular}
\caption{\label{tablelisttypes extremalRES in B}List of the types of the singular fibers of extremal rational elliptic surfaces.}
\end{table}  
\endgroup

\end{document}